\documentclass[twocolumn,twocolappendix]{aastex701}
\usepackage{graphicx}
\usepackage{amsmath, subcaption}
\usepackage{scalerel}
\usepackage{soul,xcolor}
\setstcolor{blue}

\newcommand{\um}{{\textmu}m}

\shorttitle{Highly Amorphous Silicates in 3I/ATLAS}
\shortauthors{Belyakov et al.}

\begin{document}

\title{JWST/MIRI Finds ISM-like Amorphous Silicates in Interstellar Comet 3I/ATLAS}

\correspondingauthor{Matthew~Belyakov}
\author[orcid=0000-0003-4778-6170]{Matthew Belyakov}
\affiliation{Division of Geological and Planetary Sciences, California Institute of Technology, Pasadena, CA 91125, USA}
\email{mattbel@caltech.edu}

\author[orcid=0000-0001-9665-8429]{Ian~Wong}
\affiliation{Space Telescope Science Institute, 3700 San Martin Drive, Baltimore, MD 21218, USA}
\email{iwong@stsci.edu}

\author[orcid=0000-0002-9548-1526]{Carey~M.~Lisse}
\affiliation{Johns Hopkins University Applied Physics Laboratory, Planetary Exploration Group, Space Department, 11100 Johns Hopkins Road, Laurel, MD 20723, USA}
\email{carey.lisse@jhuapl.edu}

\author[orcid=0000-0002-7451-4704]{M.~Ryleigh~Davis}
\affiliation{Department of Astronomy \& Astrophysics, University of California San Diego, La Jolla, CA 92093, USA}
\email{rdavis1@ucsd.edu} 

\author[orcid=0000-0002-4950-6323]{Bryce~T.~Bolin}
\affiliation{Eureka Scientific, Oakland, CA 94602, USA}
\email{bbolin@eurekasci.com}

\author[orcid=0000-0003-3402-1339]{Audrey Martin}
\affiliation{Division of Geological and Planetary Sciences, California Institute of Technology, Pasadena, CA 91125, USA}
\email{acmartin@caltech.edu}

\author[orcid=0000-0001-7552-1562]{Klaus~M.~Pontoppidan}
\affiliation{Jet Propulsion Laboratory, California Institute of Technology, 4800 Oak Grove Drive, Pasadena, CA 91109, USA}
\email{klaus.m.pontoppidan@jpl.nasa.gov}

\author[orcid=0000-0003-0787-1610]{Geoffrey A. Blake}
\affiliation{Division of Geological and Planetary Sciences, California Institute of Technology, Pasadena, CA 91125, USA}
\email{gab@caltech.edu}

\author[orcid=0000-0002-8382-0447]{Christine Chen}
\affiliation{Space Telescope Science Institute, 3700 San Martin Drive, Baltimore, MD 21218, USA}
\email{cchen@stsci.edu}

\author[orcid=0000-0002-8255-0545]{Michael~E.~Brown}
\affiliation{Division of Geological and Planetary Sciences, California Institute of Technology, Pasadena, CA 91125, USA}
\email{mbrown@caltech.edu}

\begin{abstract}
We present the first spectroscopic mineralogical analysis of the dust coma of an interstellar object (ISO) from JWST mid-infrared spectroscopy of 3I/ATLAS (3I). 3I exhibits a strong 9- to 11-micron emissivity feature due to Si-O vibrational stretching modes that is commonly seen in asteroids, comets, disks, and the interstellar medium. Characterization of this 10-micron emissivity maximum reveals that 3I's dust composition is dominated by amorphous silicates, and that 3I is unlike Solar System comets, which show significant crystalline silicate dust. Instead, 3I's dust composition is more similar to circumstellar transition disks and the interstellar medium. We suggest 3I may have formed in a distant part of its home system out of interstellar medium-like material, without substantial incorporation of silicates condensed near its host star, unlike the mixing scenarios commonly suggested for the Solar System. Alternatively, 3I's original crystalline silicates may have been amorphized during its Gyr-long journey, although we find this alternative less likely due to 3I's mass loss rate and distinct 10-\um\ feature as compared to observed Solar System comets.
\end{abstract}


\section{Introduction} 
Microscopic silicate dust, whether in the interstellar medium (ISM), protoplanetary disks, debris disks, or agglomerated into asteroids and comets in our own Solar System, offers one of the most useful probes into the early chemistry of planetary systems. During planetesimal formation, micron-sized dust is transformed by a series of diverse physical processes into macroscopic planetesimals \citep[][]{Simon2024Comets,Johansen2025}. The mineralogy of the observed dust traces the maximum temperature reached by embedded silicate grains \citep{gail1999mineral}, the extent of mixing between various reservoirs of silicates \citep{bockelee2002turbulent, brownleestardust}, and subsequent alteration processes such as serpentinization \citep{mcadam2015aqueous, bates2020linking}. 

Silicates in astronomical sources show notable heterogeneity, reflecting the many evolutionary stages silicate dust undergoes \citep[for a review, see][]{Henning2010ARA&A}.  Dust begins its life in outflows from asymptotic giant branch (AGB) stars or supernovae (SN), after which it can be further modified in the interstellar medium (ISM) \citep{Ferrarotti2006A&A, Gail2009ApJ, Jones2011AA}. In the ISM, dust is characteristically in amorphous, or glass-like phases, first constrained to a 5\% crystalline fraction by \cite{Li2001ApJL}, with the seminal work of \cite{Kemper2004ApJ} setting an upper limit for the amount of crystalline dust in the ISM of 2.2\% \citep[value from the erratum][]{Kemper2005ApJ}. The low fraction of crystalline dust may suggest a rapid amorphization timescale of initially crystalline material to $\sim$0.1 Gyr of ISM exposure \citep{Kemper2005ApJ, Min2007A&A}. By contrast, protoplanetary disks show a wide range of crystalline mineralogies, with forsterite/enstatite (the magnesium endmembers of olivine/pyroxene, respectively) typically seen \citep{Bouwman2008ApJ}. These crystalline silicates are a product of annealing or condensation of infalling ISM-like material near the host star \citep{Gail2004AA}. Later-stage transition disks, where planets have already formed to clear out gaps and photoevaporation has pushed the disk edge to several au, show a range of crystalline/amorphous silicate compositions \citep{Calvet2005ApJL, Espaillat2011ApJ}. Finally, debris disks, which are gas-poor with dust produced by planetary formation processes, collisional grinding of planetesimals, and cometary sublimation, show the broadest range of silicates, including typical ferromagnesian silicates \citep{Li1998AA, Chen2006ApJS, Lisse2008ApJ} along with more exotic silica species including quartz and obsidian \citep{Lisse2009ApJ}, as well as phyllosilicates \citep{Currie2011ApJ}.

Comets provide discrete snapshots of material accreted at specific times and places within our Solar System before being emplaced into the Oort cloud \citep{Duncan1987AJ}. Mid-infrared spectroscopy ($\sim$5-35 \um) of cometary dust tests the evolution of our own Solar System \citep{Hanner2010, Wooden2017RSPTA}. Following the detection of abundant crystalline silicates in comet Halley and several other bright comets in the 1980's and 1990's \citep{Ney1982come, Hanner1985AdSpR, Combes1986Natur, Campins1989ApJ, Hanner1994ApJ}, Spitzer Space Telescope comet spectroscopy revealed that crystalline silicates are ubiquitous in dusty comae \citep{LisseDeepImpact,Sitko2011AJ,Kelley2017Icar, Harker_2023}. Other notable remote sensing detections include phyllosilicates, carbonates, and iron sulfides \citep{LisseDeepImpact}, most of which have been validated by in situ sampling from the Stardust sample return mission \citep{Brownlee2014AREPS}. Stardust collected small chondrule fragments, metal oxides, and iron-nickel minerals embedded within larger grains. These results likely indicate large-scale mixing occurred in the early Solar System, transferring high-temperature condensates from the inner to outer Solar System, where they mixed with volatile ices and a small fraction of presolar or ISM grains \citep{Brownlee2014AREPS, Wooden2017RSPTA}. 

However, the transferability of knowledge accumulated from Solar System studies to planetesimal formation around systems with different stellar masses, cluster environments, and other key parameters is uncertain. In this context, interstellar objects (ISOs) can bridge the gap between observations of extrasolar disks and Solar System comets by providing direct examination of dust from another planetary system that traces an unknown yet specific formation scenario at a given time and place. While sample returns from the Stardust mission provided ground truth for remote-sensing measurements \citep{brownleestardust, LisseDeepImpact}, discrete ISOs can provide ground truth for the mineralogy of bulk dust measured in disks.

3I/ATLAS (3I) is the third interstellar object detected transiting through our Solar System and was discovered by \citet{Denneau2025MPEC} at a heliocentric distance of 4.5 au, displaying a markedly hyperbolic orbit. Preliminary characterization of its dynamics, physical properties, and composition by several groups uncovered that 3I would likely be extraordinarily bright compared to past interstellar objects 1I/'Oumuamua and 2I/Borisov by the time of its perihelion approach \citep{Bolin20253I, Seligman2025ApJL, Chandler2026ApJL}. Its $\sim$1.3 km radius nucleus \citep{Hui2026ApJL}, combined with intense outgassing of hypervolatile species such as methanol, methane, and carbon monoxide \citep{Cordiner2025ApJL, Belyakov2026ApJL, Lisse2026ApJL, Lisse2026RNAAS}, caused 3I to reach a total magnitude comparable to large Solar System periodic comets near perihelion. Given the mid-infrared capabilities of the James Webb Space Telescope (JWST), which were unavailable during the passages of the previous two interstellar objects, 3I provides a timely and potentially unique opportunity to characterize the dust preserved by an extrasolar planetesimal.

In this Letter, we follow up on the groundwork laid by our analysis of mid-infrared gas fluorescence features in 3I in \cite{Belyakov2026ApJL}, now focusing on using the mid-infrared flux continuum to obtain an emissivity spectrum of 3I. We present the first analysis of silicate mineralogy in an interstellar object, contextualizing it within the suite of available comet observations and astrophysical dust phenomena. In the discussion, we combine results on gas production, isotope ratios, and now silicate mineralogy to tell the story of 3I's formation and subsequent evolution.

\section{Observations and Data Reduction} \label{sec:obs}
3I/ATLAS was observed with the Medium-Resolution Spectrograph (MRS) on the JWST Mid-Infrared Instrument (MIRI) during UT 2025 December 15--16 and 27. The MIRI MRS instrument consists of four integral-field units (IFUs) that together cover 5--28~\um; three spectral grating settings (short, medium, long) are available, each sampling discrete subsets of the full wavelength range. Our observations cycled through all three grating settings across separate visits to obtain a continuous spectrum of 3I. Each observation was paired in sequence with analogous exposures of a nearby empty field to remove background. The full observational details are presented in \citet{Belyakov2026ApJL}, which describes 3I's mid-infrared gas fluorescence. 

Data reduction was carried out in the same manner as described by \citet{Belyakov2026ApJL}. Using the \texttt{jwstspec} tool \citep{jwstspec}, the uncalibrated detector images were processed through Version 1.20.2 of the JWST calibration pipeline \citep{jwst} to generate flat-fielded, dark-corrected, flux-calibrated, spatially-rectified, background-subtracted, and dither-combined data cubes. The non-point-source behavior of 3I's extended coma leaves us unable to leverage point-source-optimized extractions such as those of the JDISCS collaboration \citep{Pontoppidan2024ApJ}. Instead, the irradiance spectrum of 3I was extracted using a $3''$-diameter circular aperture centered on the measured centroid position. This large aperture was chosen to maximize the encircled flux and sample the full near-nucleus region. 

\begin{figure}
    \centering
    \includegraphics[width=\linewidth]{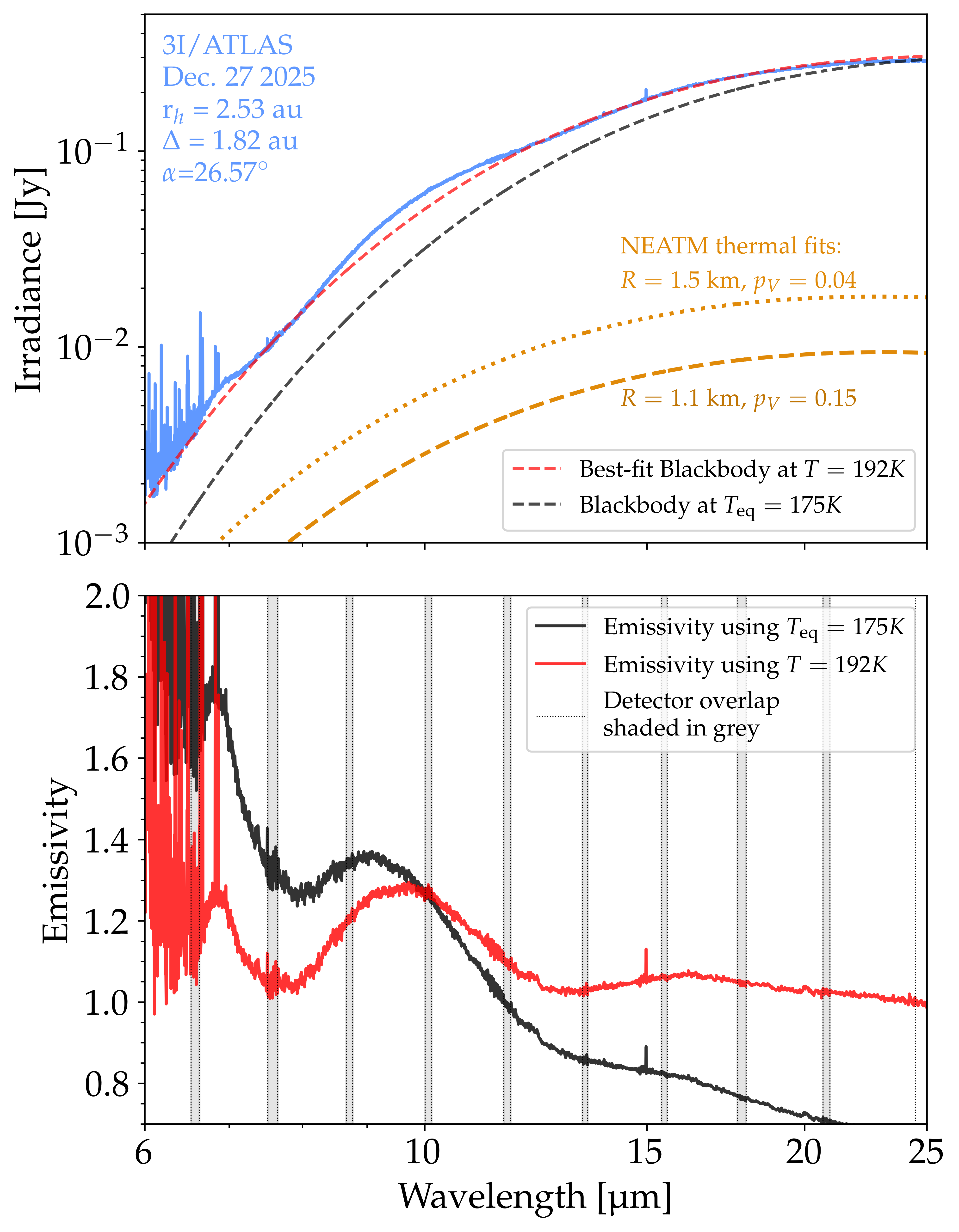}
    \includegraphics[width = \linewidth]{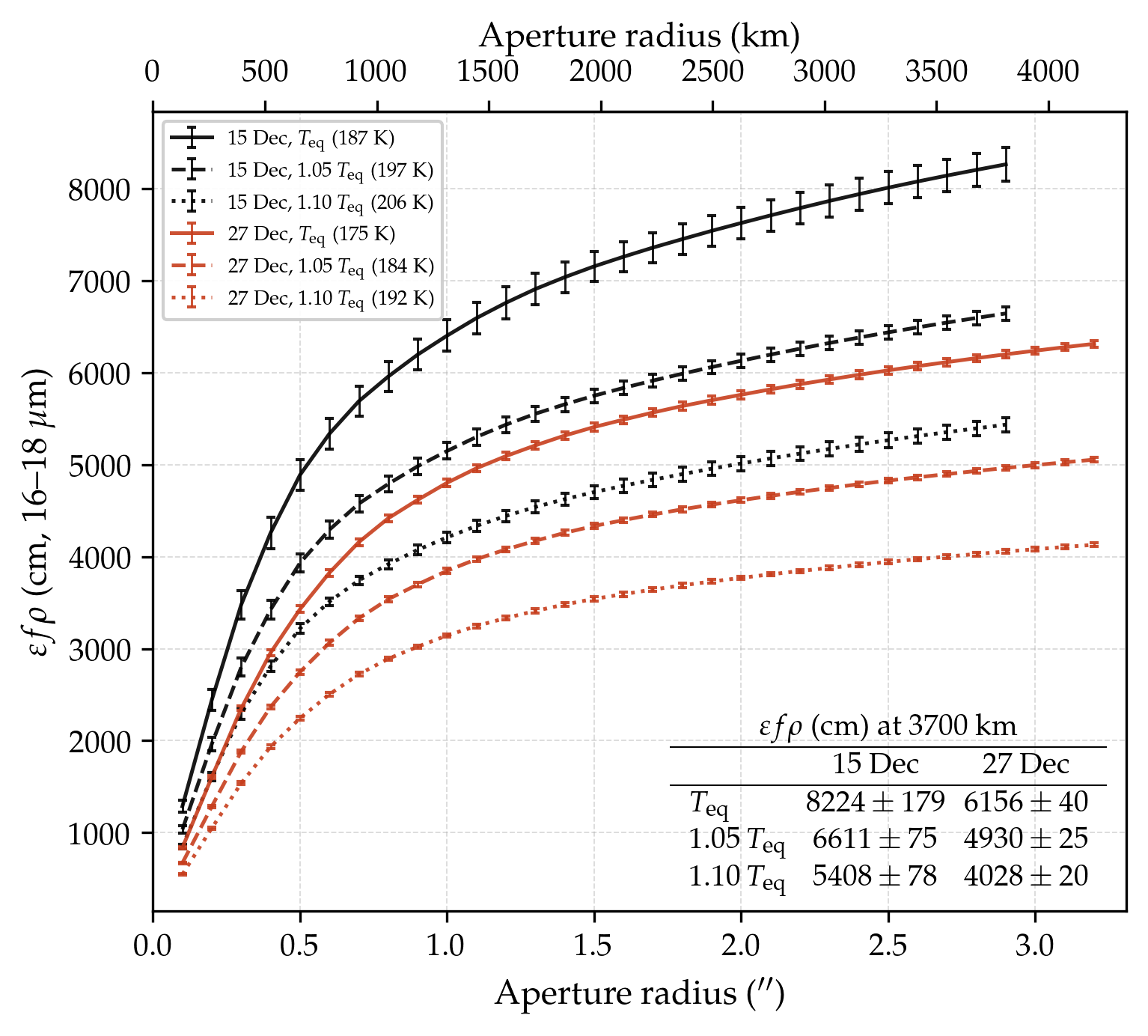}
    \caption{\textit{Top:} JWST/MIRI MRS spectrum of 3I/ATLAS in blue, overlaid with a best-fit (red) and equilibrium temperature blackbody (black). NEATM thermal fits to 3I's nucleus for two different assumed radii/albedos are shown as dotted/dashed orange curves. \\\ \textit{Middle:} Emissivity spectra of 3I/ATLAS obtained by dividing out the corresponding blackbody fits from the top panel. Shaded areas trace regions of overlap between different IFU/grating pairs.\\\ \textit{Bottom:} Aperture-averaged $\varepsilon$f$\rho$ over the 16--18\,\um\ region as a function of aperture radius on 15~December 2025 (black) and 27~December 2025 (red). Solid, dashed, and dotted lines assume blackbody temperatures $T = T_{\rm eq}$, $1.05\,T_{\rm eq}$, and $1.10\,T_{\rm eq}$, respectively. The inset table lists median $\varepsilon f\rho$ at $3700$~km for each epoch and temperature.}
    \label{fig:spec_bb}
\end{figure}

For three out of the six visits, 3I was situated very close to the edge of the MIRI MRS field of view, which precluded the use of the $3''$-diameter aperture. Our analysis, therefore, focused only on the three visits (Observations 6, 13, and 15) for which the target was well-centered. The combined irradiance spectrum is shown in \autoref{fig:spec_bb}. Adjacent segments collected across different visits were aligned by scaling the average flux levels in overlapping regions, using the measured irradiance at 9--10~\um\ (obtained on December 27 as part of Observation 15) as the absolute benchmark, since two of the three visits are from that date.

\subsection{Thermal Continuum}
Dividing mid-infrared spectra by their blackbody continuum to obtain emissivity is useful in revealing silicate features, which can be directly compared with laboratory measurements of mineralogical species. We first consider the contribution of the nucleus to the thermal flux by adopting the estimated radius of 3I of 1.3$\pm$0.2 km from \citet{Hui2026ApJL}. The radius estimate of \citet{Hui2026ApJL} assumes an albedo of 0.04 and was taken in visible light, where size and albedo are degenerate. Therefore, we estimated lower and upper limits on the flux contribution from the nucleus in a Near-Earth Asteroid Thermal Model \citep[NEATM,][]{Harris1998Icar} fit by assuming the estimate of the radius from \citet{Hui2026ApJL} and a rescaled radius estimate assuming a 0.15 albedo. Using NEATM, the nucleus's thermal contribution to the observed flux is at most 10\% and could be as low as 1\%, so most of the flux is contributed by coma dust. In this case, it is reasonable to use a single blackbody to remove continuum flux as an empirically motivated approximation of the complexity of the temperature gradient in the coma and the minor nuclear contribution. The measured irradiance spectrum, thermal fits, and emissivity spectra are shown in \autoref{fig:spec_bb}. 

We test two temperatures for continuum removal: one is the equilibrium temperature of a blackbody at 3I's heliocentric distance ($T_{\textrm{eq}} =278/\sqrt{r_{\textrm{h}}}=175$K), and one is a best-fit temperature of 192 K (also known as a color temperature $T_\textrm{C}$) to the region between 7.5-8.0 \um\ and 12.5-25.0 \um, avoiding the location of the known significant emissivity feature from silicate dust. The ``superheat'' ratio $S = T_\textrm{C}/T_\textrm{eq}$ is simultaneously affected by grain size and by composition \citep{Ney1974Icar, Gehrz1992Icar, Lisse1997EM&P}. For 3I, this ratio is $\sim$1.1, similar to many past cometary observations \citep{Lisse2002COSPA, Sitko2004ApJ, Harker_2023}. Comets typically have an $S$ between 1 and 1.3, where comets with strong anti-solar tails, prominent sub-micron dust, and prominent silicate features such as Hale-Bopp exhibit higher values of $S$ \citep{Lisse2002COSPA,Yang2009AJ}. Comets such as 2P/Encke have color temperatures near unity, which can be interpreted as indicating shallow particle-size distributions with a preponderance of $>$10~\um-sized particles \citep{Lisse2004Icar}. Photometric analysis of 3I's coma morphology is consistent with an interpretation that large grains are predominant in its coma \citep{Jewitt2025ApJL,Ren2026arXiv}. The lack of a strong anti-solar tail driven by radiation pressure is also consistent with a particle size distribution with a larger peak particle size \citep{Chandler2026ApJL}.

\subsection{Dust Production}\label{sec:dustprod}

Two related quantities have been used to estimate the dust production rate from photometry of cometary comae. The A(0$^\circ$)f$\rho$ parameter estimates the total cross-section of grains within a chosen field of view, and was introduced as a proxy for dust production \citep{AHearn1984, Fink2012Icar}. With the advent of Spitzer and numerous thermal-infrared observations of comets, a similar quantity, $\varepsilon$f$\rho$, was developed \citep{Lisse2002COSPA, Kelley2013Icar}. As is implied from the change of variables, $\varepsilon$f$\rho$ folds in emissivity instead of the particle albedo.

In the lower panel of \autoref{fig:spec_bb}, we show 3I's $\varepsilon$f$\rho$ at both observational epochs and for three assumed particle temperatures. We report the value of $\varepsilon$f$\rho$ at 16-18 \um, to have common wavelengths at both dates, and to have a measurement comparable to Spitzer's 16\um\ MIPS filter. Uncertainties are estimated using the 2-$\sigma$ scatter in $\varepsilon$f$\rho$ measured in each given wavelength slice. As noted by \cite{Kelley2013Icar} and \cite{Schambeau2021PSJ}, the quantity $\varepsilon$f$\rho$ is heavily dependent on the assumed color temperature, and may increase with aperture radius out to significant distances, as was noted for 29P/SW1 \citep{Schambeau2021PSJ}. For the December 15 observations, using a 3700 km aperture (maximum aperture entirely within the field of view), we report $\varepsilon$f$\rho=8224\pm179$ cm assuming a superheat of unity, and $\varepsilon$f$\rho=5408\pm78$ cm for $S=1.1$. Given the linearly increasing radial trend in $\varepsilon$f$\rho$ observed at these small aperture sizes, the true value is likely higher. For the December 27 observations, we report $\varepsilon$f$\rho=6156\pm40$ cm for $S=1$ and $\varepsilon$f$\rho=4028\pm20$ cm for $S=1.1$. Comparing our measurements to values for $\varepsilon$f$\rho$ from figures 4 and 5 of \cite{Bauer2017AJ}, 3I is within the upper range of observed dust production rates. However, among similarly sized comets, 3I has notably higher dust production---typically only $\sim$10~km-diameter cometary nuclei exhibit $\varepsilon$f$\rho>10^3$ cm.

We can now compare the mid-infrared $\varepsilon$f$\rho$ measurements to A(0$^\circ$)f$\rho$ from contemporaneous optical imaging obtained using Gemini (observational details described in \autoref{sec:gemini}). We calculate an r-band A(0$^\circ$)f$\rho$ = 2464.4$\pm$26.2 cm, significantly higher than the A(0$^\circ$)f$\rho$ calculated for 3I in July by \citet[][]{Bolin20253I}. Varying the apertures at radii of 1\arcsec, 2\arcsec, and 3\arcsec, we find, respectively, A(0$^\circ$)f$\rho$ = 2397.1$\pm$25.5 cm, 2396.9$\pm$25.5 cm, and 2435.4$\pm$25.9 cm. Comparing the values for the two measured dust production proxies, we find that A(0$^\circ$)f$\rho$ is between 2-4 times smaller than $\varepsilon$f$\rho$. This result is within the range of the few existing joint measurements of these two quantities (see Table 1 of \citealt{Kelley2016PASP}), and is consistent with typical assumptions regarding the emissivity, albedo, and filling factor of cometary dust.

Interestingly, $\varepsilon$f$\rho$ increases as a function of aperture size while A(0$^\circ$)f$\rho$ does not. The aperture-size dependence of $\varepsilon$f$\rho$ indicates that the coma does not follow a $1/\rho$ surface-brightness profile for particles with diameters similar to the observational wavelength. Fitting a radial profile to the MIRI/MRS data cubes finds a $1/\rho^{0.75}$ dependence instead. Thus, the coma particles do not follow ballistic expansion at some constant velocity. The shallow radial dependence of the mid-infrared surface brightness profile can be explained by fragmentation of large dust grains at increasing cometocentric distances, either from collisions or from the sublimation of water ice --- H$_2$O is known to have an extended source in 3I's coma \citep{Lisse2026ApJL}. 

\section{Dust Composition} \label{sec:silicates}
Taking the emissivity spectrum of 3I as shown in the middle panel of \autoref{fig:spec_bb}, we can begin analysis of the object's silicate composition. Before presenting a spectral model in \autoref{sec:model} based on the results of the Deep Impact and Stardust in situ Solar System comet experiments \citep{LisseDeepImpact}, we adopt a qualitative and comparative approach to interpreting 3I's spectrum. The spectral signature which enables characterization of silicate dust is the set of mid-infrared ($\sim$5-30~\um) emissivity features caused by Si-O stretching and bending modes \citep[see, for instance,][]{1968ApOpt...7...53V, Conel1969JGR, Salisbury1992RSEnv, Salisbury1992Icar, Chihara2002Pyroxenes,Koike2003Olivines,Hamilton2010ChEG,Lane2011JGRE,Speck2011Disordered}. We begin by isolating strong emissivity maxima in the spectrum and comparing them to known features in laboratory, cometary, disk, and ISM spectra. 3I shows two key emissivity features between 6 and 25 \um: a band at 6.9 \um\ and the prominent 10-\um\ emissivity feature common to various silicates.

\begin{figure*}
    \centering
    \includegraphics[width=\linewidth]{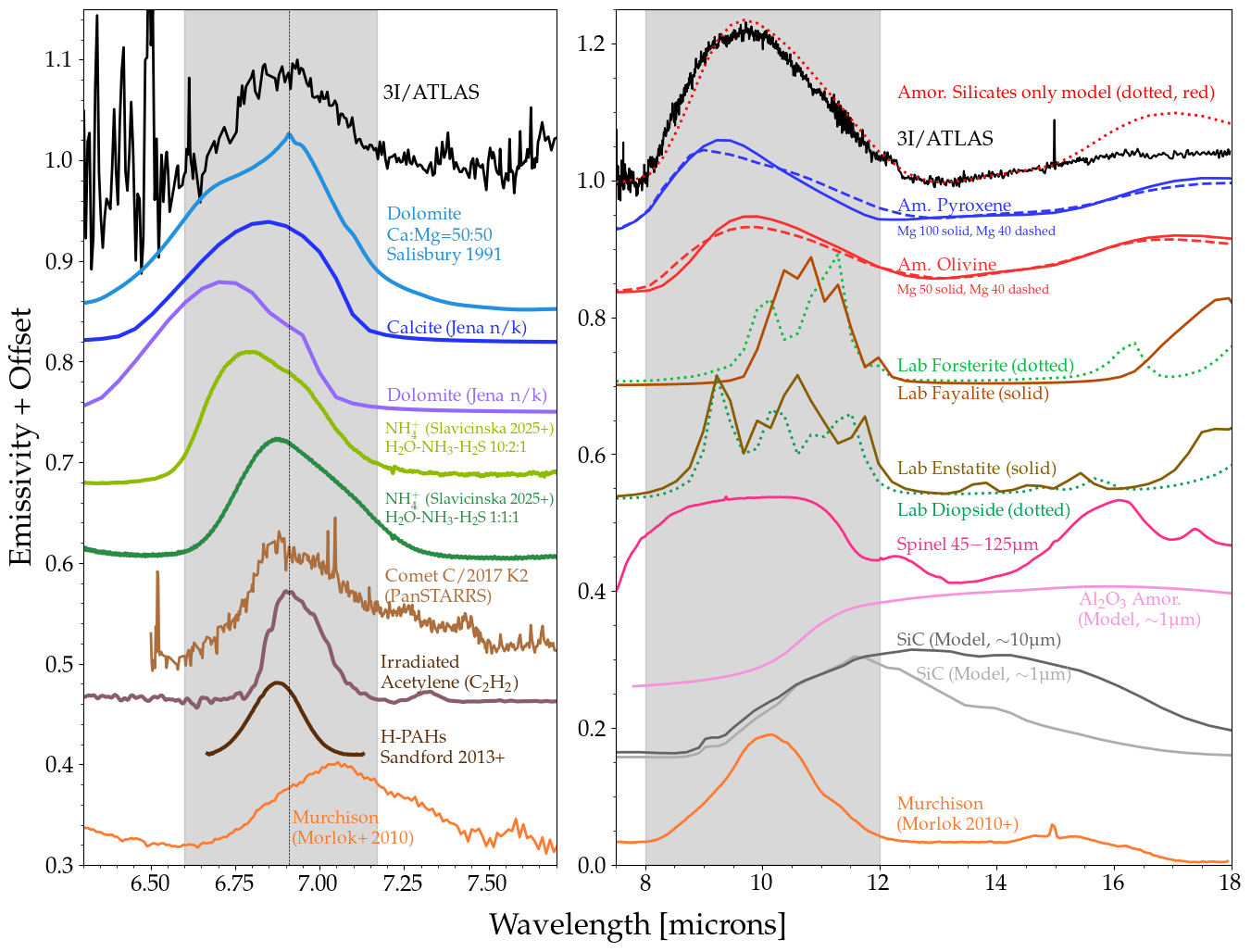}
    \caption{\textit{Left:} Continuum-subtracted emissivity of 3I/ATLAS in the 6-7 \um\ region in black, compared to a set of reference spectra which are described in the main text. The carbonates, ammonium ion, and hydrogenated amorphous carbon all match 3I's spectrum in this region. The Murchison meteorite, on the other hand, does not match the 6.9~\um\ feature in 3I, and a comparable feature from C/2017 K2 PanSTARRS (light brown) is broader and shifted to longer wavelengths \citep{Woodward2025PSJ}.
    \\ \textit{Right:} Thermal continuum-subtracted emissivity of 3I/ATLAS between 7.5-18~\um\, highlighting the 10-\um\ silicate feature. The main text describes the reference laboratory spectra. Overlaid on 3I's emissivity spectrum are two spectral models based on a simple linear combination of materials, which do not account for particle sizes or other effects. A model consisting solely of amorphous olivine and pyroxene with varying Mg/Fe ratios is shown as a red dotted line. The model fits the 10~\um\ region well, but fails at longer wavelengths. Murchison meteorite matrix material, again shown in orange, does not match 3I's 10~\um\ feature.}
    \label{fig:minerals}
\end{figure*} 

\subsection{The 6.9 \um\ feature}
The 6.9 \um\ or 6.85 \um\ feature has been discussed in the astronomical literature for decades as a ``mystery'' band, with a highly contested attribution. The first detections of this band came from the Kuiper Airborne Observatory observing massive protostars such as W33a \citep{Soifer1979ApJL, Tielens1984ApJ} or dense nebulae such as NGC 7027 \citep{Russell1977ApJL}. Observations in subsequent decades further resolved this band \citep[see, e.g.][]{Keane2001AA}, and detected it on interplanetary dust particles in our Solar System \citep{Sandford1985ApJ, Sandford1986Sci,Tomeoka1986Sci}, as well as samples of the Murchison (CM) meteorite and CI meteorites \citep{Fredriksson1988Metic}. 

In the Solar System literature, the 6.9 \um\ band is typically attributed to carbonates \citep{Knacke1980AA, Morlok2010Icar}, which are easily formed geochemically via rock-water-carbon dioxide reactions \citep{Fredriksson1988Metic}. The first detection of the mid-infrared emissivity feature of carbonates in comets was in 1P/Halley \citep{Bregman1987AA}, verified by in situ mass-spectrometer measurements by the two Vega spacecraft \citep{Fomenkova1992Sci}. Carbonates were also reported in the Deep Impact ejecta by \cite{LisseDeepImpact}. Finally, the stratospheric dust collection of comet 26P/Grigg–Skjellerup by \cite{Busemann2009EPSL} directly recovered carbonates. Whereas in meteoritic spectra and in IDPs the presence of carbonates has been directly verified, attribution in the astronomical literature varies significantly. The weaker carbonate emissivity features at 11.3 \um\ remain undetected in protostars \citep[see the discussion in][for example]{Schutte1996AA}, and a mechanism for producing carbonates without aqueous alteration in evolved planetesimals remains elusive \citep[see, however, recent work by][]{Lang2026ApJ}. The primary argument in favor of carbonates is the detection of a $\sim$ 92 \um\ feature from ISM observations of dust in protoplanetary nebulae \citep{Kemper2002AA, Kemper2002Natur}, though theoretical considerations place formation of such material in stellar wind outflows as highly unlikely \citep{Ferrarotti2005AA}. It is plausible that the 6.9 \um\ band as seen towards protostars and as detected in Solar System refractory material is a result of different carriers. Numerous non-carbonate attributions have been suggested in the astrophysical literature: calcium oxides \citep{Kimura2005ApJ}, C-H deformation in simple organic molecules such as methanol \citep{Tielens1984ApJ,Schutte1996AA}, aliphatic organic dust \citep{Greenberg1995ApJL}, and the ammonium ion \citep{Schutte2003AA}.

In the left panel of \autoref{fig:minerals}, we display the continuum-removed 6.9-\um\ emissivity maximum of the 3I spectrum, along with a sample of materials representative of the range of attributions discussed in the literature for the 6.9-\um\ feature. To each continuum-subtracted band, we fit a Gaussian to obtain a rough estimate of the band center and width to enable quantitative comparisons; 3I's 6.9 \um\ band is centered at 6.894$\pm0.004$~\um\, with a 0.31$\pm0.01$~\um\ FWHM. The Murchison meteorite (taken from \citealt{Morlok2010Icar}) appears to be a poor match to 3I as it is shifted to significantly longer wavelengths. We show three examples of carbonates: a dolomite (CaMg(CO$_3$)$_2$) laboratory reflectance spectrum (inverted using Kirchhoff's law, $\varepsilon=1-R$) taken from \cite{salisbury1991infrared}, and spectral models of ($\sim$1\um) calcite (CaCO$_3$) and dolomite generated using optical constants from the Jena database\footnote{Optical constants for calcite and dolomite at 200K can be found at \url{https://www2.astro.uni-jena.de/Laboratory/OCDB/carbonates.html}} \citep{Posch2007ApJ}. To obtain the absorption efficiency ($Q_{abs}$) from the refractive indices, we use the \texttt{glitterin} code \citep{Yurkin2011JQSRT, Lin2025PASP}, which predicts absorption efficiencies for irregular particles using a neural network approach. The calco-magnesian carbonates in \autoref{fig:minerals} roughly match the shape and appearance of the 6.9 \um\ feature on 3I. The dolomite from \cite{salisbury1991infrared} has the same band center as 3I, though it displays slightly more structure than 3I's band. The model calcite is shifted 0.03~\um\ shorter, but has the same bandwidth. Finally, the modeled dolomite peaks 0.12~\um\ shorter, and siderites and other carbonate species peak at yet shorter wavelengths compared to 3I's 6.9 \um\ feature \citep{LaneChristiansen1997, Bishop2021ESS}. Small changes in the Mg/Ca ratios, grain sizes, and temperature properties of the dust can significantly affect the appearance of the carbonate band, making a precise match difficult. A common objection to the detection of carbonates is the lack of longer wavelength features between 10-13 \um\ (see Figure 1 of \citealt{LaneChristiansen1997}). However, in laboratory-produced samples of amorphous carbonates at low temperatures, these secondary features may be weak or absent \citep{C7NR05347A,MEHTA2022121262, Gao2023Mine}. 

The NH$_4^+$ ion, appearing in the ammonium hydrosulfide salt (NH$_4$SH), plausibly matches 3I's 6.9-\um\ feature. Laboratory spectra of two distinct cryogenic codeposited mixtures of H$_2$O, NH$_3$, and H$_2$S from \cite{Slavicinska2025AA} are shown in \autoref{fig:minerals} in green. The water-dominated mixture has a 6.822 $\pm$ 0.001 \um\ band center, shorter than 3I's band, while the equal mixture of water, ammonia, and hydrogen sulfide has a 6.906 $\pm$ 0.001 \um\  band center. The ammonium ion appears to be the currently favored explanation for the 6.85 or 6.9 \um\ feature in interstellar ices \citep{Schutte2003AA,Boogert2015ARAA}, specifically in this NH$_4$SH carrier salt \citep{Slavicinska2025AA}. In 3I's coma, the particles' temperatures (170-200 K) are above or at the desorption temperature of many ammonium salts. In \cite{Slavicinska2025AA}, no features of NH$_4^+$ were visible above 174 K, and the salt crystallized between 135-155 K depending on the mixture, at which point it had a far narrower spectral signature, contrary to the band observed on 3I. Work by \cite{Loeffler2015Icar} found slightly higher temperature constraints, with significant spectral changes occurring at 180K. Other variants for the ammonium salts, such as ammonium isothiocyanate (NH$_4^+$SCN$^-$) and ammonium formate (NH$_4^+$HCOO$^-$), may be more favorable carriers of the ion because they are stable up to 230-260 K \citep{Bergner2016ApJ,ammoniumsalts}. It is unlikely that the 6.9 \um\ feature observed in 3I's dust has the same main carrier as the 6.85 \um\ feature of interstellar ices because the appearance of the main interstellar carrier correlates with water ice and likely has similar volatility. If the ammonium ion, in some undetermined carrier, is responsible for 6.85 band, we can estimate the ammonium-to-water ratio and compare it to values seen in YSOs. Using the band strength for NH$_4^+$ reported in \cite{Schutte2003AA}, the column density for ammonium is approximately (1.0 $\pm$ 0.5) $\times$ 10$^{17}$ cm$^{-2}$. This measured column density is over an order of magnitude smaller than the contemporaneous water column density in \cite{Belyakov2026ApJL}. The ammonium-to-water ratio is thus consistent with expectations from measurements in protostars \citep{Oberg2011ApJ, Slavicinska2025AA}.

We now turn to the literature on organic compounds for further comparisons. Polycyclic aromatic hydrocarbons (PAHs) and other organics have been confirmed in \textit{Stardust} samples \citep{Clemett2010MPS}. Given 3I's abundance in CH$_4$, C$_2$, and other organic molecules, it behooves us to examine this possible attribution for the 6.9 \um\ band. In the analysis of the comet C/2017 K2 (PanSTARRS), \cite{Woodward2025PSJ} suggest hydrogenated PAHs or long-chain aliphatic organics as an explanation for a very similar 6.9 \um\ band observed on C/2017 K2. The attribution of this feature to PAHs requires explaining why the canonical 6.2, 7.7, 8.7, 11.3, and 12.7 \um\ bands are absent or not apparent at the signal-to-noise level of the observations. For the 6.9 \um\ band to appear as the sole dominant feature from organic compounds, either all of the PAHs must be fully hydrogenated (H-PAHs), or the contribution must be dominated by only aliphatic organics, as articulated by \cite{Sandford2013ApJS}. We show the spectrum of H-PAHs taken from \cite{Sandford2013ApJS} in the left panel of \autoref{fig:minerals}. The H-PAH band is half as wide as the one seen in 3I, but has a similar band center (0.02 \um\ short). While H-PAHs could plausibly contribute to the 6.9 \um\ feature, studies of PAH-rich ISM regions do not find the 6.9 \um\ band in isolation \citep{Sloan1999ApJL, VanDePutte2025A&A} -- even the most hydrogenated-PAH regions have a stronger 6.2 \um\ signal than at 6.9\um\ \citep{Materese2017ApJ}. Thus, PAHs are unlikely to be responsible for 3I's 6.9 \um\ feature.

Refractory aliphatic hydrocarbon dust and hydrogenated amorphous carbon (HAC) offer another possible explanation for the 6.9 \um\ feature. We show a spectrum of EUV-irradiated C$_2$H$_2$ in the left panel of \autoref{fig:minerals}. The experimental data are preliminary results from \cite{lee2026euv}, with the full publication in preparation. Irradiated acetylene shows an asymmetric 6.9 \um\ feature, with a second peak at 7.33 \um. The primary 6.9 \um\ feature is narrower than in 3I, while the secondary feature is short of 3I's 7.36 \um\ peak. Though not a precise match, irradiated acetylene is, among the organic materials, the closest to 3I's spectrum, as it lacks the numerous PAH features that are conspicuously absent on 3I. In HACs this secondary feature is attributed to the CH$_3$ symmetric bending mode and is centered at 7.273 \um\, clearly short of 3I's 7.36 \um\ feature \citep{Gadallah2012AA}. In the band assignments for irradiated organic compounds from Table 2 of \cite{Hand2012JGRE}, no band matches that of 3I's 7.36 \um\ feature. We discuss the attribution of the 6.9 \um\ feature further in \autoref{sec:69dontget}.

\begin{figure*}
    \centering
    \includegraphics[width=\linewidth]{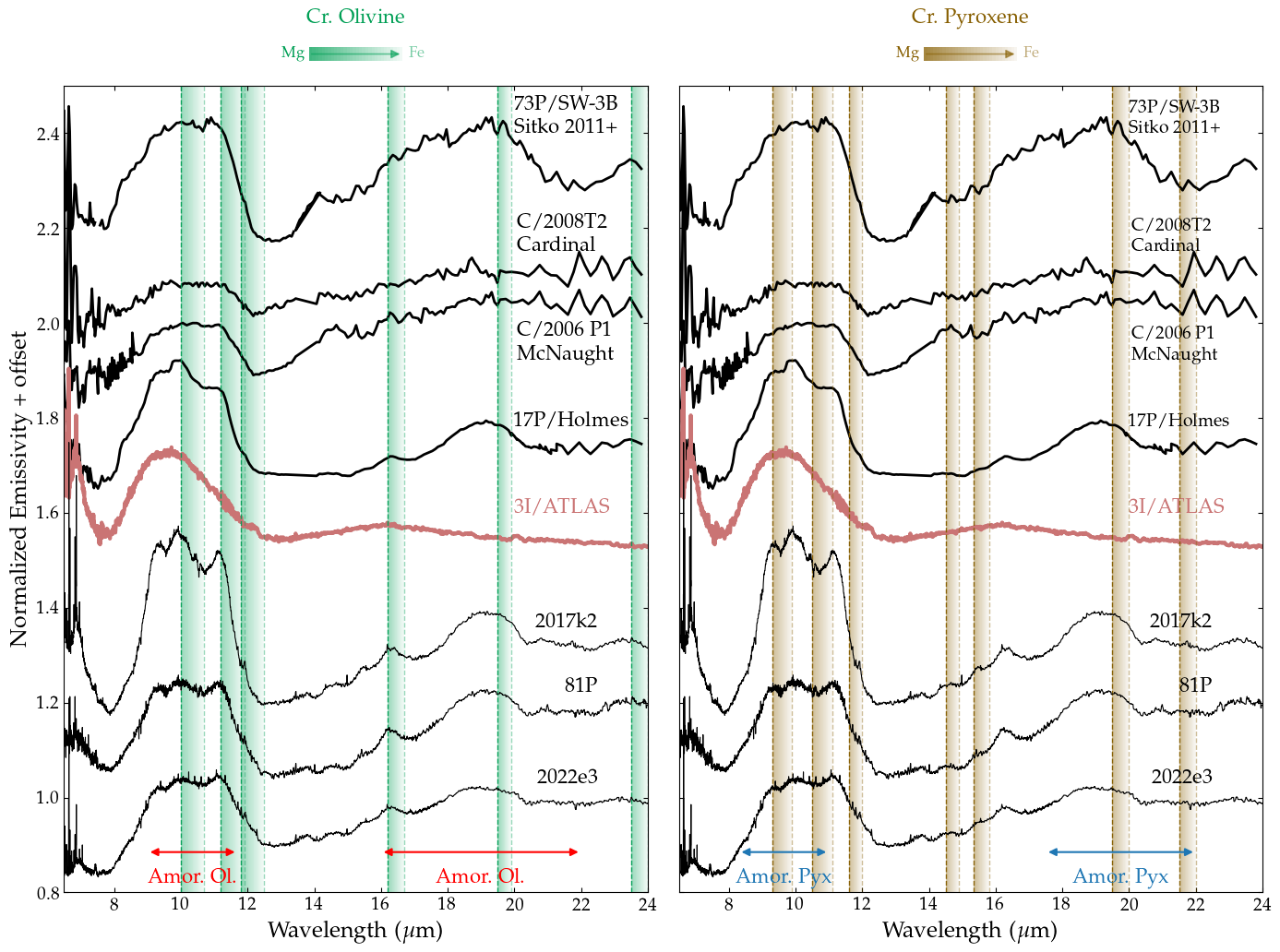}
    \caption{Comparison of 3I to Solar System comets observed by Spitzer and JWST, with the location of crystalline olivine/pyroxene (left/right panels) peaks highlighted as shaded bands which cover the range of possible peak values for possible Fe/Mg ratios. The arrows at the bottom of the panels show the locations of the amorphous olivine and pyroxene emissivity maxima. C/2017 K2 was previously presented in \cite{Woodward2025PSJ}, 17P/Holmes (originally studied by \citealt{Reach2010Icar}) and C/2006 P1 McNaught (originally studied by \citealt{Lisse2007CBET}), and C/2008 T2 Cardinal are taken from \cite{Harker_2023}, while 73P/SW-3B is taken from the version published in \cite{Sitko2011AJ}. 3I is unique because it shows no obvious crystalline silicate peaks. Comets with similar 18 \um\ contrasts, and perhaps similarly large grain sizes, still evince crystalline features at long wavelengths. C/2006 P1 McNaught is perhaps the closest analog to 3I, yet it also has some clear structure from crystalline silicates.}
    \label{fig:comets}
\end{figure*}

\subsection{The 10-\um\ Feature}
3I shows the strong 10-\um\ feature typical of comets \citep{Hanner1985AdSpR,Hanner1994ApJ, Hanner1997EMP,Lisse2007Icar,Kelley2009PSS,Harker_2023}. The 10-\um\ feature is attributed to the Si-O stretching mode in silicate minerals, and is highly diagnostic of the specific mineralogy and metal ratios in the minerals, and is sensitive to particle size \citep{salisbury1991infrared, Mustard1997Icar, Martin2025JGRE}. Longer-wavelength features can provide additional information that can measure these quantities. 3I's continuum-subtracted 10-\um\ feature is shown in black in the right panel of \autoref{fig:minerals} along with a set of possible contributors to the 8-18~\um\ spectral region.

We show models of amorphous pyroxene and olivine of various Mg/Fe ratios in blue and red, respectively. These are generated using optical constants from the Jena database \citep{Jaeger1994AA, Henning1997AA, Jager2003AA} and the aforementioned \texttt{glitterin} code, assuming a size parameter $2\pi\cdot a$/$\lambda$ of unity. Even a cursory examination of 3I's composition reveals that the dust is dominated by a mixture of amorphous silicates, as there is no significant structure in the 10~\um\ region other than a rounded peak with a maximum around 9.5 \um. A red dotted line overlaid on 3I's spectrum shows a best-fit linear combination of these amorphous silicates (mostly 50/50 Mg/Fe olivine) that matches 3I's 10~\um\ emissivity maximum, though not the region past 15 \um. However, the fact that a simple model can reproduce the key feature in 3I's spectrum well enough without a physically motivated model hints that amorphous silicates are the predominant contributors to the dust's composition. 

Examining the crystalline Mg-rich and Fe-rich endmembers, which are taken from \cite{Chihara2002Pyroxenes} and \cite{Koike2003Olivines}, there are a significant number of narrow peaks not obviously visible in 3I's emissivity spectrum, though they could be present as minor constituents. There is a slight inflection in the spectrum of 3I at 16.1-16.3 \um\ which may be explained by some small fraction of crystalline silicates, specifically forsteritic olivine. A narrow feature at 12.2 \um\ is also seen in 3I, however, this wavelength corresponds to the location of the known spectral leak in MIRI/MRS \citep{Gasman2023AA}. Although we applied the spectral leak correction step from the JWST pipeline, the feature appears to match the known FWHM and central location of the leak, which the JWST pipeline may not fully correct. 

Metal oxides, especially calcium-aluminum species, arise as common high-temperature condensates in protoplanetary disks and are a probable ISM constituent. Small, potentially pre-solar grains of spinel have been found in meteorites \citep{Zega2014GeCoA} as well as Stardust samples \citep{lerouxstardust}, and may be responsible for a 12-13 \um\ feature in AGB stars \citep{Fabian2001AAspinel}. We show a lab-measured spinel (provenance described in \autoref{sec:spinel}) in the right panel of \autoref{fig:minerals} in pink -- the spinel shows a broad 10-\um\ feature with a plateau unlike 3I, as well as clear structure at longer wavelengths. We also show a model of fine-grained amorphous alumina generated from Jena optical constants \citep{Begemann1997ApJ} and the \texttt{glitterin} code using a size parameter of 0.1 corresponding to 1 \um\ grains -- larger grains produce a featureless spectrum. Both of these aluminum oxides appear unlikely to be major contributors to 3I's dust.

Where ferromagnesian silicates and metal oxides are thought to arise from O-rich/C-poor condensation, silicon carbide is thought to commonly form in high C/O environments \citep{Zhukovska2008AA}. Numerous works have discussed the C/O ratio in 3I through the lens of CO/CO$_2$ production \citep{Lazzarin_2026, Lisse2026ApJL}, though this measurement primarily corresponds to the availability of ices and temperature at the formation location of the object, rather than the initial stellar C/O ratio, which is probed by the ratio of condensed silicates to carbides \citep[see, e.g.][]{Shakespeare2025AJ}. SiC has famously been found in meteorites \citep{Bernatowicz1987Natur} and is reported in outflows from carbon-rich stars \citep{Speck1997MNRAS}. Notably, direct detection remains difficult in the ISM \citep{Whittet1990MNRAS}, which may be due to oxidation in the ISM or a result of grain size effects preventing clear detection \citep{Chen2022MNRAS}. We use optical constants from the Jena database \citep{Mutschke1999AA} at two different size parameters (0.1 in light grey, 1 in dark grey) to generate the SiC spectra shown in the right panel of \autoref{fig:minerals}. SiC does not appear to be a necessary component of 3I's spectrum. At small grain sizes, which are thought to be dominant specifically for silicon carbide \citep{Chen2022MNRAS}, the SiC feature is narrow and does not appear to contribute to 3I's spectrum. For large grain sizes, SiC would not produce the right slope at longer wavelengths because of a 16 \um\ downturn. 

Finally, we compare 3I to an entirely different class of material -- matrix material from the Murchison meteorite. The 10-\um\ band in Murchison shown in orange (taken from \citealt{Morlok2010Icar}) comes from Fe-Mg serpentine, a phyllosilicate that forms from aqueous alteration and is common in meteorite matrix material. The serpentine feature is narrower than 3I's 10-\um\ feature and is shifted to longer wavelengths. Though, past 12 \um, the Murchison matrix is spectrally neutral, not unlike 3I. While it is possible that serpentine is a minor contributor to 3I's spectrum, it is not likely given the lack of evidence for other altered phases commonly seen in cometary material.

\subsection{Comparisons to Comets from Spitzer and JWST}

Comets are, at present, the only widely available tracer of the mineralogical history of material in the outer Solar System. The combination of decades of mid-infrared observations \citep[e.g.][]{Hanner1994ApJ} with the in situ measurements from the numerous Halley, \textit{Deep Impact,} and \textit{Stardust} missions have revealed that comets comprise a mixture of inner Solar System material --- crystalline silicates, phyllosilicates, and carbonates --- and outer Solar System material possibly inherited from the local ISM \citep{bockelee2002turbulent, Brownlee2014AREPS}. In this picture, comets provide evidence for large-scale mixing in the protoplanetary disk \citep{bockelee2002turbulent, LisseDeepImpact, Brownlee2014AREPS}. Our analysis thus far suggests that 3I exhibits a highly amorphous silicate composition, possibly dissimilar to Solar System comets. Therefore, we examine the available mid-infrared cometary spectra from Spitzer and JWST to evaluate the extent of the irregularity of 3I's spectrum relative to Solar System comets.

Three comets have been observed at high SNR with MIRI MRS prior to 3I: 81P \citep{Roth2026arXiv81p}, C/2022 E3 (ZTF) \citep{Foster2026AJ}, and C/2017 K2 (PanSTARRS) \citep{Woodward2025PSJ}. We reduce these observations following our method for 3I. Where 3I, 81P, and C/2022 E3 (ZTF) have similar spectral contrasts -- ratios of the 10-\um\ peak to nearby continuum -- the latter two comets show numerous secondary peaks at longer wavelengths. The bands at 9.2, 10.0, 11.2, 14.5, and 16.2 \um\ suggest that significant proportions of crystalline silicates are present in the coma dust of 81P and C/2022 E3. C/2017 K2 shows crystalline silicate peaks that are stronger than those of the former two objects and is very dissimilar to 3I. The only obvious crystalline feature that may appear in both the sample of JWST MIRI MRS comets and 3I is a weak olivine peak at 16.2 \um. 

\begin{figure*}
    \centering
    \includegraphics[width=0.7\linewidth]{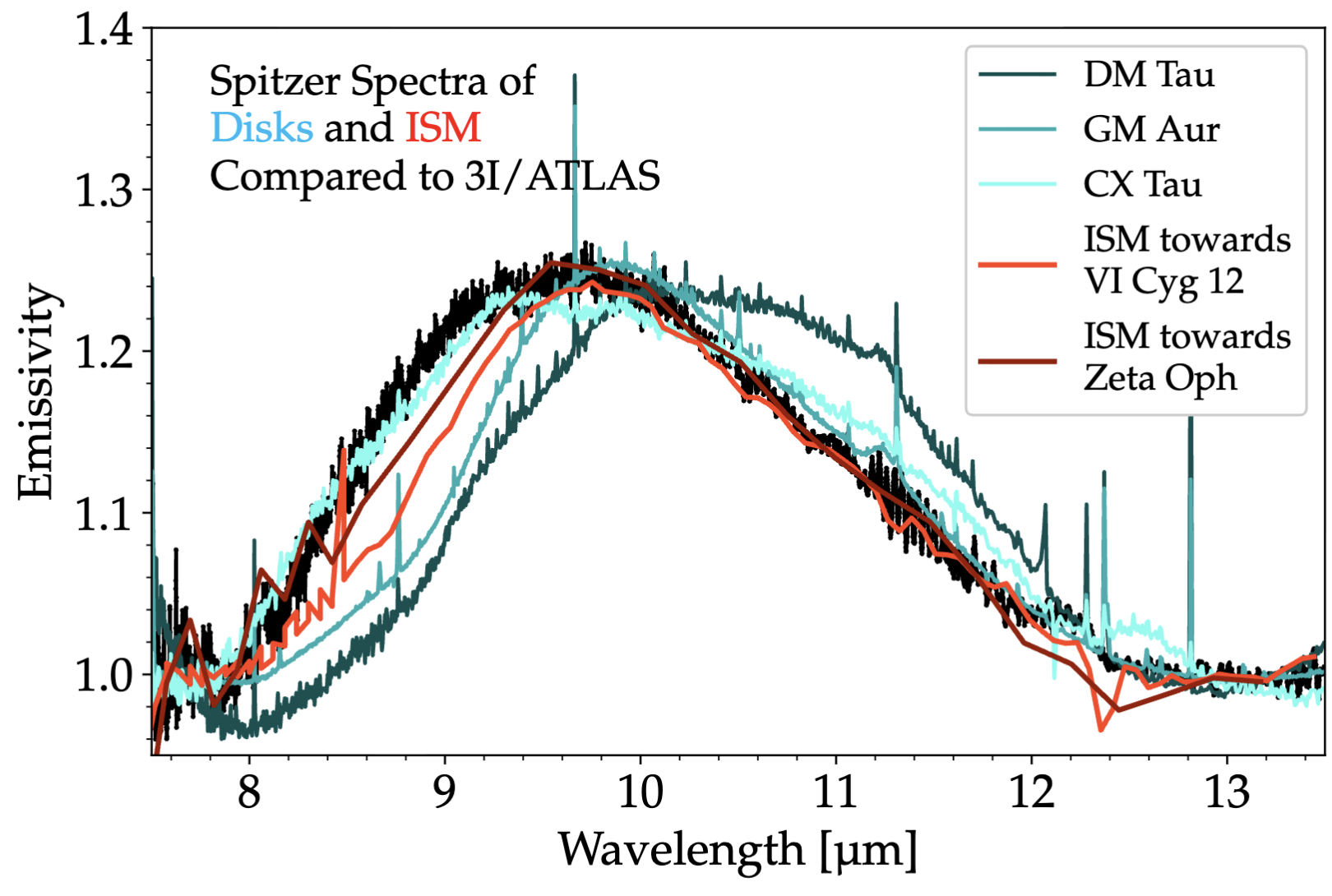}
    \caption{Comparison between 3I/ATLAS, several transition disks, and ISM lines-of-sight. The three disks (blue/cyan lines) show a range of pyroxene/olivine and Mg/Fe ratios, with CX Tau representing the best match to 3I. ISM silicates, especially as seen towards $\zeta$ Oph, largely match 3I's 10-\um\ feature, though they do not fully capture the shorter-wavelength slope.}
    \label{fig:disksnism}
\end{figure*}

Turning to the larger Spitzer sample (spectra taken from \citealt{Harker_2023} and \cite{Sitko2011AJ}), 73P/SW-3 was a disrupted comet (and thus revealing untouched subsurface material) observed by Spitzer that appeared to be notably poor in crystalline silicates \citep{Sitko2011AJ}. Nonetheless, all fragments of 73P/SW-3 show a trapezoidal 10-\um\ peak, unlike the rounded, triangular feature on 3I, and they have longer-wavelength structure. 17P/Holmes, notable for its major outburst and high dust production rate, differs from 3I. Highly active, Sun-skirting comet C/2006 P1 McNaught has low spectral contrast \citep{Kelley2007AAS, Lisse2007CBET}, and may be the closest match to 3I. C/2006 P1 McNaught appears to lack a significant fraction of crystalline silicates, showing a more rounded feature than its counterparts. Even so, its 10-\um\ feature is closer to the flat-topped, trapezoidal 10-\um\ shape of 81P than it is to 3I. 

In the spectral modeling that \cite{Harker_2023} presents for the Spitzer comet sample, several comets are reported to have 0\% crystalline fractions\footnote{Notably, a fragment of 73P is reported as having 99\% crystalline fraction, which \cite{Harker_2023} suggests is a non-physical outlier. A 0\% crystalline fraction may be similarly suspect without inspection of individual spectra.}. When comparing 3I to the entire ensemble of comets presented in \cite{Harker_2023}, we do not find a single comet that resembles 3I's remarkably rounded 10~\um\ feature, which is a precise match to ISM amorphous silicates, as we discuss in \autoref{sec:compareism}. C/2008 T2 Cardinal, also shown in \autoref{fig:comets}, has the most statistically significant detection of a 0\% crystalline fraction in the work of \cite{Harker_2023}. The 10-\um\ feature of C/2008 T2 Cardinal is of notably low spectral contrast, yet still shows a trapezoidal-like 10-\um\ feature. We note that this trapezoidal feature is due to a sharp inflection at 11.2 which suggests the presence of at least some crystalline olivine. While a given effort at modeling can construct 0\% crystalline fractions for 3I, some Solar System comets, or both, 3I's 10-\um\ feature is evidently dissimilar to the sample of over 60 Solar System comets observed at mid-infrared wavelengths.

\subsection{Comparisons to Astrophysical Sources}
\label{sec:compareism}
We now turn to comparing 3I/ATLAS to disks and material from the interstellar medium, as surveyed Solar System comets fail to provide an adequate analogue to 3I. We obtained JWST MIRI/MRS observations of disks using the JDISCS pipeline \citep{Pontoppidan2024ApJ}, including GM Aur, DM Tau, and CX Tau; the latter two are from Program ID 1282, \citealt{Henning2017jwst}; GM Aur is from Program ID 2025, \citealt{Oberg2021jwst}). We also retrieved Spitzer observations of ISM clouds towards bright sources: $\zeta$ Oph and VI Cyg 12 \citep{Poteet2015ApJ, Fogerty2016ApJ}. The continuum-removed 10-\um\ regions of these five sources are overlaid on 3I's spectrum in \autoref{fig:disksnism}.

DM Tau and GM Aur are transition disks around T Tauri stars; Spitzer analyses of their dust are presented in \cite{Calvet2005ApJL} and \cite{Sargent2006ApJ}. Transition disks such as DM Tau and GM Aur are at the end of their lifetimes and as such show significant structure, such as rings of millimeter-sized dust. This structure is hypothesized to result from photoevaporation of the inner disk edge, with gaps developing further out due to giant-planet formation or strong pressure bumps from ice lines \citep[for a review, see][]{Bae2023}. High spatial resolution observations at radio wavelengths by ALMA have demonstrated that the disks around GM Aur and DM Tau have relatively narrow rings of millimeter-sized dust at dozens of au away from the host star, with large gaps hosting optically thin dust in between  \citep{Huang2020ApJ, Hashimoto2021ApJ}. In these transition disks, the observed dust at 10-\um\ is predominantly amorphous. This composition is attributed to the inner disk being cleared and thus not contributing to the total observed spectrum of the disk. In these disks, the outer part does not receive significant influx of material from the inner part of the system \citep{vanBoekel2004Natur}. While the compositions of these disks are modeled as amorphous-only (with more pyroxene in DM Tau than GM Aur), the 10-\um\ feature in 3I is at shorter wavelengths than in these disks, either due to distinct particle size distributions, pyroxene/olivine ratios, or Mg/Fe numbers in the silicates.

CX Tau, a low-mass T Tauri star, provides the best match to 3I of the disks we examined from the Spitzer archive, and was modeled by \cite{Sargent2009ApJS} as strongly dominated by amorphous olivine. ALMA observations of CX Tau indicate that the disk lacks mm-sized dust beyond 20 au, while the gas disk extends much farther \citep{Facchini2019AA}. The observations do not preclude abundant small dust in this distant region -- the difference in PSD may be responsible for pushing the 10-\um\ feature to shorter wavelengths. Inspection of the JWST spectrum of CX Tau in Figure 1 of \cite{Vlasblom2025AA} reveals that a small amount of crystalline olivine is likely also present, which may explain the difference in the spectrum of CX Tau from 3I at 11 \um\ in \autoref{fig:disksnism}.

Finally, ISM material captures most of the behavior of 3I's 10-\um\ feature. Spectra of the ISM taken through two distinct sightlines are shown in \autoref{fig:disksnism} from \cite{Fogerty2016ApJ}. These spectra were modeled using a mixed stoichiometry Mg-rich ``polivine'', or Mg$_{1.5}$SiO$_{3.5}$, and have a relatively close match to 3I's spectrum. \cite{Fogerty2016ApJ} apply the same model to the aforementioned DM Tau and GM Aur disks, finding that introducing polivine does not improve their fits -- the olivine component drives the shorter wavelength peak of the ISM sightlines compared to transition disks. However, differences in grain-size distributions may explain this effect just as well, and these two astronomical phenomena may have very different distributions. In either case, it is clear that the appearance of 3I's 10-\um\ feature is well-approximated by a number of astrophysical sources rich in amorphous dust; Solar System comets do not provide similarly good comparisons to 3I.

\subsubsection{Crystalline Fraction}

Given the numerous \textit{qualitative} indications that 3I's spectrum is dominated by amorphous rather than crystalline silicates, we provide a \textit{quantitative} estimate of the fraction of crystalline silicates in the coma dust as is commonly done in the disk and ISM literature \citep[e.g.][]{Li2001ApJL,Kemper2004ApJ, vanBoekel2005AA}. Examining the 10-\um\ band alone (7.5 to 13.5 \um), we can replicate the method of \cite{Kemper2004ApJ} to obtain a crystalline fraction. We use the same optical constants of amorphous pyroxene and olivine (\cite{Jaeger1994AA} and \cite{Dorschner1995AA}), while crystalline silicate optical constants are taken from  \cite{Fabian2001AA} for fayalite and \cite{Zeidler2015ApJ} for enstatite and forsterite. We begin by computing a mass absorption coefficient $\kappa(\lambda)$ for each species, but rather than using a continuous distribution of ellipsoids method, we use \texttt{glitterin} to obtain $C_\text{abs}$ (the absorption cross section) to obtain $\kappa = C_\text{abs}/\rho V$. The mass absorption coefficients are then continuum-subtracted (similarly to 3I in \autoref{fig:minerals}). We first fit the observed emissivity with the amorphous component only, and then increase the crystalline component -- forsterite, enstatite, and fayalite, assuming a 1.5\% calibration error added in quadrature to pipeline errors. We minimize our $\chi^2$ when we introduce a 0.4\% crystalline fraction, with a 3-$\sigma$ upper limit of 3.1\% for the crystalline fraction, dominated by the addition of forsterite. This upper limit is similar to the 3-$\sigma$ upper limit for the ISM crystalline fraction \citep{Kemper2004ApJ, Kemper2005ApJ}. However, it is far lower than typical crystalline fractions in Solar System comets (measurements by \cite{Harker_2023} suggest 25\% is typical, though model results vary widely between groups).

\section{Deep Impact Spectral Model Applied to 3I/ATLAS} \label{sec:model}
We have established that 3I's dust shows a dearth of crystalline silicates by comparing its spectrum with laboratory materials, comets, disks, and the ISM. Instead, 3I's dust is composed of amorphous or glassy silicates, along with some significant organic component evidenced by a 6.9 \um\ feature, potentially attributed to carbonates or large aliphatic chains. We now implement the Deep Impact Spectral Compositional Dust Model as described in \citet{LisseDeepImpact}, hereafter the DI model. We apply the DI model to explore whether a list of over 100 solid-state constituents typical of Solar System comets, meteorites, debris disks, and the ISM can robustly characterize 3I. Model components include calco-ferro-magnesian silicates, amorphous carbon, sulfides, water ice or gas, PAHs, and calco-ferro-magnesian carbonates. We also examine the residuals of this fit to determine whether significant additional components must be added to fit 3I's dust composition. A key point is that our implementation of the DI model combines pipeline errors in quadrature with a 1.5\% absolute calibration error, which should capture errors from mixing two observation dates with different temperatures and flux-calibration changes as a function of wavelength. Additionally, for the DI model, we mask the region between 11.5-12.5 \um --- see the discussion on issues with flux calibration at those wavelengths in \cite{Chen2024ApJ}.

\begin{figure}
    \centering
    \includegraphics[width=\linewidth]{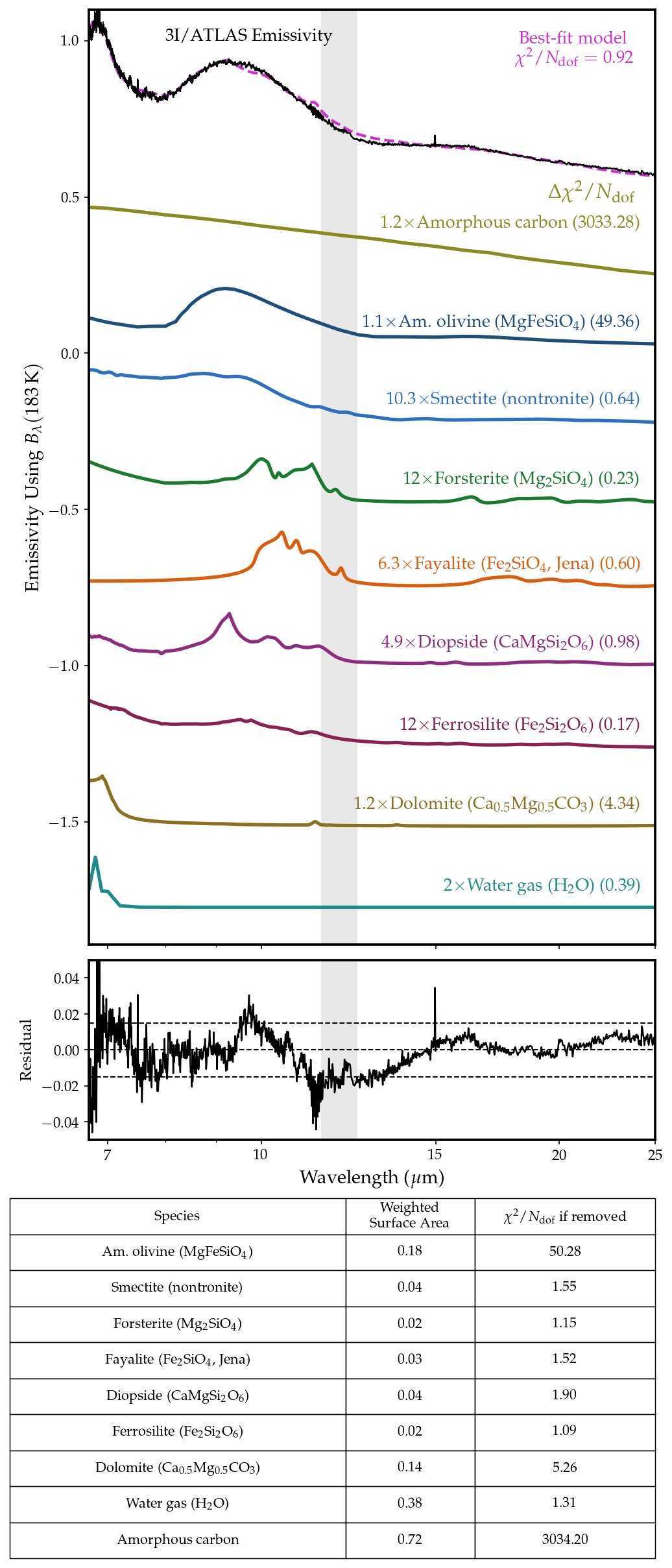}
    \caption{Application of the Deep Impact Spectral Dust Model of \cite{LisseDeepImpact} to 3I/ATLAS. Top panel shows 3I's emissivity spectrum with the best-fit model overlaid (dashed line). The grey region shows the 11.5-- 12.5 \um\ region excluded from the fit due to calibration issues. Rescaled model components with non-zero contributions are shown below the spectrum, alongside their $\Delta\chi^2/N_{\text{d.o.f.}}$ if removed from the model. Middle panel shows model residuals. The reported $\chi^2/N_{\text{d.o.f.}}$ assumes a 1.5\% systematic error added in quadrature to pipeline-derived errors, shown as dashed lines in the residual. The table lists all relevant components, their weighted surface-area contribution, and their $\Delta\chi^2/N_{\text{d.o.f.}}$.}
    \label{fig:casey}
\end{figure}

The DI model finds a rough effective temperature for the observed coma dust of $\sim$183K, only 7K higher (superheat $=1.046$) than the expected LTE temperature of 3I on December 27, 2025. The power-law particle size distribution (PSD) that best fits the spectrum is rather shallow, with d$n$/d$a\sim a^{-3.0}$, where $a$ is the particle size, compared with the typical collisional PSD power-law index of $-3.5$. The shallow size distribution derived from the DI model is consistent with other evidence pointing to a relative overabundance of large $\sim10-100$ \um\ particles, including 3I's lack of an anti-solar tail dominated by micron-sized dust undergoing radiation pressure in optical data \citep{Jewitt2025ApJL, Hui2026ApJL, Chandler2026ApJL}, as well as 3I's low color temperature.

We show the resulting model, model components, and residuals in \autoref{fig:casey}. The reported reduced $\chi^2$ is 0.92 for an assumed absolute calibration error of 1.5\%\footnote{Our 1.5\% systematic is a smaller error than has been used in past modeling efforts of mid-infrared spectroscopy. 2\% is typical and motivated by the quality of available optical constants as well as wavelength-dependent flux-calibration uncertainties}. Taking $\chi^2/N_{\text{d.o.f.}}=0.92$ as a baseline, we can report the contribution of various model components by comparing the best-fit $\chi^2$ with the values obtained when individual components are excluded. Removing amorphous carbon would yield $\chi^2/N_{\text{d.o.f.}}= 3034.2$ and removing amorphous olivine yields $\chi^2/N_{\text{d.o.f.}} = 50.3$. Removing the crystalline silicate end-members ferrosilite and forsterite reduces the quality of fit by only $\Delta\chi^2/N_{\text{d.o.f.}}\lesssim0.25$ and are within the 3-$\sigma$ limit for 1259 degrees of freedom. Water gas is significant, with $\Delta\chi^2/N_{\text{d.o.f.}}=0.39$, and it is known to be present in this wavelength range from our previous work \citep{Belyakov2026ApJL}. At slightly higher significance are fayalite (Fe crystalline silicate), smectite (phyllosilicate), and diopside (calco-magnesian crystalline silicate), with $\Delta\chi^2/N_{\text{d.o.f.}}$ of 0.6, 0.64, and 0.98, respectively, placing them at well over 3-$\sigma$ detections. Smectite mostly contributes continuum, and the detection represents the uncertainty present in modeling a spectrum with muted features due to a particle-size distribution that leans towards large particles. If smectite is present, then it may have a similar origin to the carbonates. The carbonate dolomite is the primary contributor to the 6.9 \um\ feature in the model, and has a $\Delta\chi^2/N_{\text{d.o.f.}}\sim4$, higher than the rest of the minor components combined. All told, the highest confident DI-model detections are amorphous carbon, amorphous olivine, and dolomite; crystalline silicates and hydrated minerals are plausible minor constituents.

What components are absent? Amorphous pyroxene is conspicuously missing, with no statistically significant contribution necessary. A lack of pyroxene is unusual, as past studies typically find roughly equal stoichiometric mixtures of olivine and pyroxene in cometary dust \citep{Lisse2007Icar, Lisse2012ApJ, Harker_2023}. Additionally, PAHs and other organics, such as the aforementioned irradiated acetylene, do not appear to improve the fit, primarily due to their numerous features between 7-12 \um\ which cannot be resolved in 3I's spectrum. Other solid-state species found in past observations of comets and circumstellar disks, such as metal sulfides, numerous crystalline silicates, or metal oxides such as alumina or spinel, are not detected on 3I at the level of statistical significance by the DI model. The lack of 25-50~\um\ spectra, where many of these compounds, such as metal sulfides, have emissivity maxima, may preclude our ability to definitively detect them.

While the crystalline silicates are not as prominent as the amorphous component, the DI model suggests a non-zero crystalline fraction. Therefore, we must address the DI model's putative detection of diopside and fayalite. As an upper bound, if all four crystalline species that improve the reduced $\chi^2$ are included, then the crystalline fraction is 33\% based on the reported weighted surface area fraction. Dropping the two less significant crystalline members yields a 24\% crystalline fraction. If crystalline silicates are removed from the fit entirely, a model including only carbonates, amorphous silicates, amorphous carbon, and water gas worsens the reduced $\chi^2$ to 2.22, corresponding to a 30-$\sigma$ discrepancy or a 33\% underestimation of the systematic uncertainty. 

The DI result is in contention with our previous demonstration that comets with 10~\um\ contrasts similar to 3I all evince crystalline silicates and have distinct 10-\um\ features. Moreover, the entire 13-24~\um\ region in 3I shows no notable features other than an inflection at 16~\um, despite rather good signal-to-noise. Finally, in the slope between 10-11 \um, we see no significant deviations from a straight-line slope, as in amorphous olivine. Quantifying this, if we model the narrow crystalline silicate peaks in the 9-11~\um\ region as Gaussians above a continuum, we can fit them to the data to test whether any crystalline peak is detected at a significant level. Except for the 9.3~\um\ feature of diopside detected at 4-$\sigma$, which happens to be the strongest detection of crystalline silicates in the DI model, we do not detect any crystalline peaks above the 2-$\sigma$ level, including those of forsterite or fayalite. The crystalline fraction is then 15\% if we only include diopside, a reasonable estimate given it is the only crystalline silicate with a $\Delta\chi^2/N_{\text{d.o.f.}}>$1. We note that the most significant crystalline silicate has Mg and Ca cations, and the best-fitting carbonate is also a Ca/Mg carbonate. A particle size distribution weighted towards large particles will tend to dilute any crystalline feature above the continuum, and 3I appears to have a coma dominated by large particles based on several independent lines of evidence. Thus, a 15\% crystalline fraction could be hidden within the spectrum given large enough particles. Another possible explanation is that the available optical constants poorly model actual grains, though employing new optical constants from \cite{Demyk2022AA} does not improve our fit.

Examining the residuals in the bottom panel of \autoref{fig:casey}, we find that there is a relatively narrow feature at 9.8\um\ which is not quite explained by the model, and that the shape of the 6.9 \um\ feature is not fully captured by the carbonates, which can similarly be seen from \autoref{fig:minerals}. A small abundance of carbonates, combined with some contribution from the narrower band profile of hydrogenated PAHs or aliphatic organics shown in the left panel of \autoref{fig:minerals} would be able to decrease the residuals in that region. The poor fit in the 10~\um\ range could reflect a lack of optical constants with the precise combination of stoichiometry, Fe/Mg ratio, and grain shape that matches 3I. We note that a large contributor to increasing the $\chi^2$ is the 6.9~\um\ region. Ignoring any data short of 7.5~\um\ for the purposes of estimating goodness-of-fit, the reduced $\chi^2$ drops to 0.73 when we only compute points past 7.5~\um. This large change in the effective $\chi^2$ when restricting the fitted wavelengths emphasizes our incomplete understanding of spectral features between 4.8-8~\um.

\section{Discussion} \label{sec:discussion}
With 3I now exiting our Solar System, key results on its dust composition, isotopic ratios, and its evolution during its close passage, can be combined to put together a possible story for the origin of this interstellar object. Measurements of the D/H, $^{12}$C/$^{13}$C, and $^{14}$N/$^{15}$N ratios have been obtained using a variety of species at UV-vis, infrared, and radio wavelengths. The D/H ratio in 3I has been measured in water at infrared wavelengths \citep[][IR]{Cordiner2026arXiv}, in water at radio wavelengths \citep[][]{SalazarManzano2026NatAs} and in methane in the infrared \citep[][]{Roth2026arXiv}. Although the measured D/H ratios are statistically inconsistent across these three measurements at the many-sigma level, all investigations find significant enhancement relative to Solar System comets. The $^{12}$C-$^{13}$C ratio was measured in both CO$_2$ and CO by \cite{Cordiner2026arXiv}, and in CN in the near-ultraviolet by \cite{Opitom2026arXiv}. Both studies found a notable lack of $^{13}$C or a high $^{12}$C-$^{13}$C ratio, higher than in Solar System comets. $^{13}$C is thought to be formed from secondary processing in novae and giant stars \citep{Langer1990ApJ}. Thus, the $^{12}$C-$^{13}$C ratio is expected to decrease (increase in the amount of $^{13}$C) over the lifetime of the galaxy -- our Solar System records a 4.5 Gyr old ratio which is above that of the present-day ISM.

This work has characterized the dust inventory of 3I. The key result is that 3I appears to lack a significant fraction of crystalline silicate material and instead resembles ISM dust or dust from certain late-stage transition disks. Two possible explanations emerge. The first is that the observed dust, traveling through the ISM, has experienced the same processing as ISM dust and thus been similarly amorphized. The second is that 3I formed without incorporating significant amounts of crystalline silicates. We first argue why we disfavor the former hypothesis on surface processing before discussing the implications for 3I's formation scenario.

\subsection{Do post-perihelion observations probe pristine material?}

Post-perihelion spectroscopy of 3I suggests that by December, a subsurface reservoir of unprocessed or less processed material had been emerging from the comet. Our previous work in \cite{Belyakov2026ApJL} found a distinct change in the methane production rate post-perihelion, while the work of \cite{Zhao2026arXiv} has shown that the early depletion of 3I in carbon-chain molecules evidenced from a high CN/C$_2$ ratio had changed to an even CN/C$_2$ ratio after the perihelion approach. SPHEREx post-perihelion observations by \cite{Lisse2026RNAAS} also find notable changes in 3I's volatile composition as compared to August 2025. Calculations by \cite{Frincke2026arXiv} on the volatile loss of 3I throughout its passage support the loss of at least 10 meters of surface material. 

Using our measured dust production rate, we can estimate the total mass lost by 3I and convert it to an implied change in radius. Depending on the assumed temperature, our calculated $\varepsilon$f$\rho$ for 3I on December 15 ranges between 5410 and 8220 cm, and between 4030 and 6150 cm on December 27, using a 2.7'' aperture\footnote{This aperture size is smaller than the extraction, as for the $\varepsilon$f$\rho$ calculation, we use the observation from December 15th at 15 \um\, which is not used to construct the spectrum referenced throughout the text.}. We note that the true $\varepsilon$f$\rho$ is higher based on the linearly increasing trend in the bottom panel of \autoref{fig:spec_bb}. Using these points, the dust production post-perihelion follows, roughly, a $r_\textrm{h}^{-2.25}$ power law with respect to distance from the Sun, which can be modified to a $r_\textrm{h}^{-2}$ power law to follow a physically motivated scaling proportional to incident solar flux, thus allowing us to extrapolate dust production as a function of heliocentric distance. We can then obtain a mass loss rate (derived in \autoref{sec:dustprodcalc}) using
\begin{equation} \label{eqn:dust_prod}
\dot{M}=\frac{4\pi}{3}\frac{a_{\rm eff}\rho_d v_d}{\varepsilon}\left(\varepsilon f \rho\right),
\end{equation}
where $a_\textrm{eff}$ is the effective particle size resultant from the adopted power law particle size distribution, $\rho_\textrm{d}$ is the particle density of 1 g cm$^{-3}$, and $v_d$ is the outflow velocity for dust, taken to be 50 m/s (based on Figure 1 of \citealt{Fink2012Icar}), and emissivity $\varepsilon$ is taken to be 0.9. Assuming a power-law of $q=-3$ for the particle size distribution for particles between 1 \um\ and 1 mm yields a mass-weighted $a_\textrm{eff} = 1.45\cdot 10^{-4}$ m. However, if we assume particles as large as 1 cm appear, our production rates are 7.5 times larger. From the range of the two $\varepsilon$f$\rho$ values from December 15, we obtain an instantaneous dust production rate between 1820 and 2770 kg/s.

We normalized a $r_h^{-2}$ mass loss rate to the $\varepsilon$f$\rho$ value measured on 2025 December 15 using Equation 1, and then integrated it from 2025 June 1 through 2025 December 27, using the $r_h$ values for 3I from JPL HORIZONS. Then, assuming a spherical nucleus, the final radius is
\begin{equation}
R_f = \left(R_0^3 - \frac{3M_{\rm loss}}{4\pi\rho_n}\right)^{1/3},
\end{equation}
where $R_0$ and $\rho_n$ are the initial radius and bulk density of the nucleus -- we assume a radius range of $R_0$=1.1-1.5 km and $\rho_n=0.5~{\rm g~cm^{-3}}$. Our calculation is based upon a slew of assumptions, but taking the worst-case scenario of a large nuclear radius, particles no larger than 1 mm, and a superheat for the dust of 1.1, we obtain an erosion of 2.5 meters. On the other hand, assuming a smaller radius and centimeter-sized particles at $T_\textrm{eq}$, we can obtain a radius loss as high as 56.5 m. This range of radius loss corresponds to a total object size diminution between 0.2 and 5\%. 

Our post-perihelion observations of 3I's dust or its isotopic ratios may reflect the interstellar object's natal heritage, assuming optimistic parameter choices for the particle size and other parameters. However, at the lower end of the radial range (2.5-10 m), 3I's observed dust composition may simply reflect the same ISM processing that causes initially crystalline silicates to become amorphous. Further constraints on the penetration depth of cosmic rays and other processing, such as SN shocks, are needed to determine whether the surface exposed at the time of our observations represents pristine material shielded from possible sources of alteration \citep{Maggiolo2026ApJL}. Additionally, whether galactic cosmic rays (GCRs) or SN shocks cause silicate amorphization may change the answer to the nature-versus-nurture question for 3I's silicate composition \citep[see, for example, the discussion in][]{Jones2011AA}. We emphasize that our calculations represent a lower bound to possible mass loss. We do not account for the mass loss from gas as discussed at length by \cite{Frincke2026arXiv}, who find a volatile-only radial diminution of 1-6 m. Our model does not account for any outbursts during perihelion when 3I was unobservable from Earth. Additionally, non-uniform mass loss, such as from a breach in an otherwise altered surface, provides other opportunities for 3I to release unaltered material.

While 3I's radial change does not conclusively constrain the relevance of surface processing to the observed composition, the comparison to comets strengthens our argument that we are seeing unprocessed material. Oort cloud comets in the Solar System spend the majority of their lifetimes well outside of the heliopause, yet even with 4.5 Gyr of processing, even weakly-active and dynamically new Oort cloud comets show significant fractions of crystalline material in their comae \citep{LisseDeepImpact,Harker_2023}. Oort cloud comets experience the full brunt of processing as they travel, in effect, through the ISM, even if bound to the Sun. As has been suggested in previous work \citep{Jones2000JGR,Jones2011AA}, GCRs are unlikely to be the driver of amorphization of crystalline silicates -- and it is unlikely that the flux of GCRs on Solar System comets or 3I is radically different. However, SN shocks, which are more stochastic, would have had to affect 3I far more strongly to explain the observed silicate composition. Based on the available evidence we proceed with the interpretation that 3I's measured spectral properties reflect inherited material from its birth system.   

\subsection{Attribution of the 6.9 \um\ feature}\label{sec:69dontget}

The carrier of the 6.9 \um\ feature in 3I's spectrum remains unidentified. Comparison of the feature with lab materials finds that carbonates such as dolomite provide the best match. The lack of any other strong features from PAHs or organics in the infrared, except a 3.4-3.5 \um\ feature observed by JWST NIRSpec and SPHEREx, which can be at least partially explained by methanol, methane, and H$_2$CO \citep{Lisse2026RNAAS,Roth2026arXiv}, seems to disfavor most organics as the sole explanation for the 6.9 \um\ feature in 3I. Only aliphatic hydrocarbon dust can account for 3I's strong 3.4 \um\ and 6.9 \um\ features without producing numerous other significant bands \citep{Greenberg1995ApJL}. Even hydrogenated amorphous carbon \citep[see, e.g.][]{Dartois2007AA} has a secondary feature at 7.25 \um\ not present in 3I's spectrum, though HACs are a much closer match to 3I's possible organic component than aromatic organics. Attributing 3I's dust to refractory aliphatic organics requires an explanation for why 3I lacks the aromatic hydrocarbons common to both ISM dust \citep{Li2012ApJL, Yang2017NewAR} and comets \citep{Schuhmann2019AA}.

If carbonates are responsible for the 6.9~\um\ feature, they need not be a dominant component, as even small amounts result in a very strong emissivity maximum at 6.9~\um. Objections referencing that carbonates display longer-wavelength features at 11-14~\um\ \citep{LaneChristiansen1997} are countered by studies finding that amorphous calcium carbonate has significantly weaker secondary bands \citep{andersen1991infrared, Gao2023Mine}. Additionally, perhaps the carbonate ion itself, rather than carbonate minerals, could be producing the feature. Geochemically, carbonate is found wherever CO$_2$ and water are in proximity to silicate minerals -- the water and CO$_2$ gas flowing out of the comet in the near-nucleus region, combined with the dust, may be sufficient to produce the carbonate feature. Recent laboratory work has begun to hint at this formation pathway \citep{Rouille2020ApJ, Lang2026ApJ}.

A more important objection to carbonates is that their anhydrous formation had not been convincingly demonstrated around protostars \citep{Ferrarotti2005AA}. However, recent laboratory experiments imitating the growth of refractory matter on interstellar grains have formed carbonates from embedding organic matter onto silicate grains \citep{Rouille2020ApJ, Lang2026ApJ}. In situ formation of the carbonate ion in carbonated silicate grains should also be considered \citep[see, for example][who generate a range of features in the 6-8 \um\ range from ablation of silicate grains with carbon dioxide, water vapor, and O$_2$ present]{Rouille2024ApJ}.

If, instead, the ammonium ion is the molecule responsible for the 6.9 \um\ emission, our results would be confirmation of the results of past work indicating that ammonium salts are prevalent in the ISM and are a primary sink for nitrogen in interstellar ices \citep[for example][]{Schutte2003AA, Slavicinska2025AA}. Such an identification would also reinforce the idea that 3I inherited primarily unprocessed interstellar material. However, the $\sim180-200$K temperatures of 3I's particles are rather high, and many carrier salts of the ammonium ion are expected to crystallize or desorb at these temperatures \citep{Loeffler2015Icar,Slavicinska2025AA}, giving features rather unlike the broad asymmetric Gaussian we observe (see left panel of \autoref{fig:minerals}). We suggest that some combination of carbonates and refractory aliphatic hydrocarbons may plausibly explain the 6.9 \um\ feature, whereas ammonium salts appear unlikely given the dust temperature.

\subsection{Where and how did 3I form?}
Based on the available evidence, we outline some possible constraints and options for 3I's formation. As discussed in several recent works on the isotopic composition of 3I \citep[see][]{Cordiner2026arXiv, Opitom2026arXiv, SalazarManzano2026NatAs}, the ices in 3I could not have equilibrated with nebular H$_2$ gas given the significantly elevated D/H ratio. A low $^{13}$C abundance may pin 3I's formation to early times in galactic history, linked to $^{13}$C production in novae and giant stars. Our results suggest that 3I formed by incorporating ISM ices without recondensation and without significant admixture of silicates transported outward through its parent disk.

Crystalline silicates form from condensation in a disk close to a host star or subsequent annealing of amorphous solids at high temperature \citep{Kemper2004ApJ, Gail2009ApJ} -- 3I avoided incorporation of silicates that had undergone these processes, and instead formed out of amorphous phases, likely inherited directly from the ISM. Solar System comets are endowed with a significant fraction of crystalline solids as evidenced by crystalline-rich dust found by Stardust and the Deep Impact experiments. 3I's mid-infrared spectrum is unlike any Solar System comet observed, and instead resembles observations of transition disks and ISM sightlines, reflecting a formation driven by incorporation of predominantly ISM material. Evolutionary models of dust mineralogy suggest that disks reach some degree of crystallinity early in their evolution, based on interpretations of observed disk surfaces \citep[see, e.g.][]{Oliveira2011ApJ}. Close to the host star ($\sim$1 au), dust will condense and crystallize on timescales of less than 1 Myr \citep{Ciesla2007Sci}. Thus, 3I could have formed either distantly and early enough to avoid mixing, or in a parent system with strong barriers to outward transport of crystalline silicate material. 

One possible formation scenario that suppresses radial mixing is a dead zone in 3I's parent system, as described in \citet{Charnoz2019AA}. This would occur in a disk's midplane, where magneto-rotational instabilities are suppressed due to the lack of ionized gas, creating an ohmic dead zone. Even in this low-turbulence environment, small overdensities form, enabling planetesimal formation \citep{Yang2018ApJ}. A dead zone at a significant distance in a parent disk could potentially satisfy the isotopic requirements and silicate requirements simultaneously. Alternatively, the theoretical simulations of \citet{Andama2024AA} suggest that quiescent disks do not form planetesimals efficiently at the water ice line, but that planetesimal formation may be enhanced in high-metallicity quiescent disks at distant secondary ice lines. Formation at the CO$_2$, CO, or CH$_4$ ice lines would help explain 3I's significant enhancement in these compounds relative to H$_2$O \citep{Belyakov2026ApJL, Lisse2026RNAAS}. Modeling 3I's formation may be challenging, as dust traps in structures created by planetary formation have recently been shown to be leaky \citep{Stammler2023AA, VanClepper2025ApJ, Houge2026AA}. Additionally, measuring the spatial dependence of the crystalline/amorphous ratio in the silicate feature in debris disks (e.g. \citealt{Weinberger2003ApJL, Lu2022ApJ}) may help determine whether such scenarios are plausible.

These conclusions are contingent upon interpreting 3I's composition as tracing the material the object accreted with, rather than subsequently altered dust. However, the combination of 3I's mass loss, changing volatile production regime, and lack of analogous Solar System comets leads us to disfavor the alteration-driven hypothesis for 3I's amorphous silicates. Given our story for 3I, models of disk evolution and planetesimal formation should admit a broader range of possibilities than those revealed by observations of Solar System small bodies. When considering the evolution of protoplanetary disks, large-scale radial mixing of condensed materials need not be a universal feature of planetary systems \citep{bockelee2002turbulent, Williams2011ARAA,Testi2014prpl}. Alternatively, 3I may suggest that distant planetesimals form before the onset of significant radial mixing. As more interstellar comets are studied at mid-infrared wavelengths by JWST, the range of possible conditions of planetesimal formation with direct observational support may continue to expand. 

\section{Summary}
We briefly summarize the key results of our campaign to observe 3I/ATLAS in the mid-infrared, as presented in this Letter and our previous Letter on the mid-infrared gas features of 3I/ATLAS \citep{Belyakov2026ApJL}.

\begin{itemize}
    \item A processed surface layer was removed from 3I during its perihelion passage. The evidence is twofold. First, the volatile production regime has shifted to one increasingly dominated by hypervolatiles, presumably from a subsurface reservoir. Second, estimates of the object's radial diminution from its dust production rate suggest significant mass loss.
    \item 3I's particle-size distribution is skewed towards large particles. The shallow radial dependence of 3I's mid-infrared dust production, the extended production of water gas, and the low superheat as measured from the dust's thermal emission provide evidence for an abundance of large grains in 3I's coma.
    \item 3I is rich in amorphous silicates and has a dearth of crystalline material. The emissivity spectrum of 3I lacks incontrovertible evidence for most of the sharp-peaked emissivity features common to forsterite, enstatite, or any of the other crystalline species commonly reported in comets and many disks. Instead, 3I's silicates are more similar to those observed in transition disks and the ISM. 
    \item 3I provides evidence for planetesimal formation in conditions more primitive than anything found in our Solar System. Where abundant crystalline silicates in Solar System comets suggest large-scale radial mixing, the lack of these materials in 3I's coma suggests a cold and distant origin, or a very early ejection epoch from 3I's home system. An alternative conclusion could be that 3I's crystalline silicates were destroyed in its passage through the interstellar medium; whether amorphization occurs under tens of meters of surface dust has not been investigated.
    \item Detection of the 6.9 \um\ feature on 3I continues the half-century-long mystery of the provenance of this feature in a wide variety of astrophysical phenomena. Based on the band shape, center, and temperature of 3I's dust, the likeliest attributions for the 6.9 \um\ feature are carbonates and refractory aliphatic dust.
\end{itemize}

\begin{acknowledgments}
This work is based on observations made with the NASA/ESA/CSA James Webb Space Telescope. The data were obtained from the Mikulski Archive for Space Telescopes at the Space Telescope Science Institute, which is operated by the Association of Universities for Research in Astronomy, Inc., under NASA contract NAS 5-03127 for JWST. The JWST/MIRI observations are associated with Program \#9442. The specific observations analyzed are available at \dataset[doi:10.17909/p4qv-4p68]{https://doi.org/10.17909/p4qv-4p68}. Part of this research was carried out at the Jet Propulsion Laboratory, California Institute of Technology, under a contract with the National Aeronautics and Space Administration (80NM0018D0004). We thank Alycia Weinberger, Zhe-Yu Daniel Lin, and Katherine de Kleer for enlightening discussions on mid-infrared spectroscopy. We thank Katie Slavicinska for providing laboratory spectra of NH$_4$SH and assisting in the discussion of the ammonium ion carrier. We extend a special thanks to Davide Farnocchia and Marco Micheli for updating the ephemerides of 3I, which enabled these observations. This research benefited from B. T. Bolin's participation in the 2026 workshop ``Exploring Planetary Systems in the Era of Time-domain Astronomy'', hosted by the Institute for Astronomy at the University of Hawai’i and supported by award 2106927 from the National Science Foundation’s Astronomy \& Astrophysics Research Grants program.

\end{acknowledgments}

\facilities{JWST/MIRI.}
\software{\texttt{astropy} \citep{astropy2013,astropy2018,astropy2022}, \texttt{jwst} \citep{jwst}, \texttt{jwstspec} \citep{jwstspec}, \texttt{matplotlib} \citep{matplotlib}, \texttt{numpy} \citep{numpy}, \texttt{scipy} \citep{scipy}.}

\bibliography{main}{}
\bibliographystyle{aasjournal}

\appendix

\section{Gemini Imaging}
\label{sec:gemini}
\begin{figure}[h]
\centering
    \includegraphics[width = 0.9\linewidth]{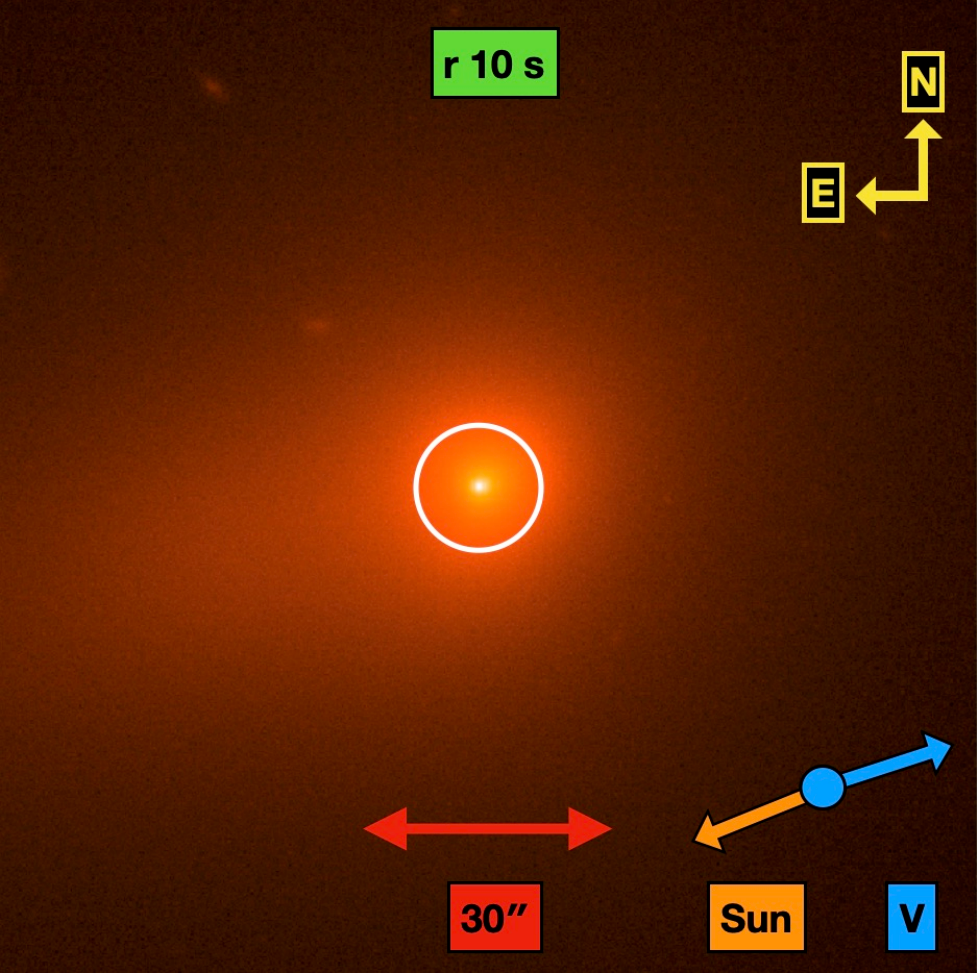}
    \caption{A 10 s r-band image of 3I taken with Gemini North/GMOS on 2025 Dec 22 12:10:37. The cardinal directions, image scale, extended target-to-Sun radius vector, and the positive of the target's heliocentric velocity vector are indicated. A white circle is shown indicating the size of the aperture used to measure the photometry using a 7.65\arcsec circular radius (10,000 km at the 1.802 au geocentric distance of the comet on 2025 December 22).\\}
    \label{fig:gemini}
\end{figure}

3I was observed using the Frederick C. Gillett Gemini 8.1-m Telescope at Gemini North on 2025 Dec 22 12:10:37 under Director's Discretionary Time program GN-2025B-DD-102 (PI: Bolin). 3I was located at RA, Dec = 10:30:43.6, +08:22:48, and was moving with an apparent sky-plane motion of 0.22\arcsec/min, with a position angle of 274$^\circ$. Fig.~\ref {fig:gemini} shows an image of 3I taken with the Gemini Multi-Object Spectrograph (GMOS) instrument \citep[][]{Hook2004} in a 10-s r-band exposure. The telescope was tracked at the comet's sky motion rate during the observations. The GMOS focal plane consists of a Hamamatsu array with a 5.5\arcmin~$\times$ 5.5\arcmin~ field of view and an effective pixel scale of 0.08\arcsec. We used the camera in 2 $\times$ 2 binning mode, giving an effective pixel scale of 0.16\arcsec. The r-band filter used is equivalent to a Sloan Digital Sky Survey (SDSS) r filter with an effective wavelength of 630 nm \citep[][]{Fukugita1996}. We measured seeing at the time of the observations using in-field stars with a FWHM of 0.56\arcsec. We took the observations at an airmass of 1.35.

Gemini photometry of 3I was calibrated using the Pan-STARRS source catalogue \citep[][]{Tonry2012,Chambers2016} following the methodology of \citet[][]{Moreno2017,Bolin2025KY26,Bolin2025YR4} for using in-field sources for photometry calibration. The photometry of 3I was measured in the Gemini r-band data using a 47-pixel radius, equivalent to a 7.65\arcsec~aperture, which had a physical size of 10,000 km at the comet's 1.8 au geocentric distance. A sky-median subtraction annulus with an inner and outer radius of 33\arcsec~and 36\arcsec was used for background removal. Portions of the sky containing the comet's tail were masked from this measurement. We measured the comet's r-band magnitude as 13.38$\pm$0.01. We calculated the A(0$^\circ$)f$\rho$ parameter at several different radii as a rough proxy for dust production rate normalized to a phase angle of 0 degrees \citep[][]{AHearn1984, Schleicher1998} -- these values are reported in \autoref{sec:dustprod}.
 % r-band brightnesses of 15.62$\pm$0.01, 15.11$\pm$0.01, 14.9$\pm$0.01, and 14.41$\pm$0.01 corresponding to
\section{Dust Production from {$\varepsilon \lowercase{f} \rho$}}
\label{sec:dustprodcalc}
To estimate 3I's dust production rate, we translate $\varepsilon f \rho$ to a mass production rate. We begin by determining the cross-sectional area of dust within the photometric aperture,
\begin{equation}
\begin{split}
\varepsilon f \rho &= \frac{\varepsilon C_x}{\pi \rho},\\
C_x &= \frac{\pi \rho}{\varepsilon}\left(\varepsilon f \rho\right).
\end{split}
\end{equation}
where $C_x$ is the geometric cross section of particles within an aperture of radius $\rho$ and for an emissivity $\varepsilon$. Instead of assuming a single particle size, we apply a particle size distribution of the form
\begin{equation}
n(a)\mathrm{d}a = \kappa a^{q}\mathrm{d}a,
\end{equation}
where $a$ is the particle radius and $\kappa$ is a normalization constant. The geometric cross section is
\begin{equation}
C_x = \int\limits_{a_{\min}}^{a_{\max}} \pi a^2 n(a)\,\mathrm{d}a,
\end{equation}
which for a power-law of $q=-3$ becomes
\begin{equation}
\begin{split}
C_x &= \pi \kappa \int\limits_{a_{\min}}^{a_{\max}} a^{-1}\,\mathrm{d}a =\pi \kappa \ln\left(\frac{a_{\max}}{a_{\min}}\right),\\
\kappa &= \frac{C_x}{\pi \ln(a_{\max}/a_{\min})}.
\end{split}
\end{equation}
The total dust mass within the aperture is
\begin{equation}
M_d= \int\limits_{a_{\min}}^{a_{\max}} \frac{4}{3}\pi a^3 \rho_d n(a)\,\mathrm{d}a, 
\end{equation}
where $\rho_d$ is the density of coma particles. Substituting the size distribution,
\begin{equation}
\begin{split}
M_d &= \frac{4}{3}\pi \rho_d \kappa \int\limits_{a_{\min}}^{a_{\max}} \,\mathrm{d}a,\\
M_d&= \frac{4}{3}\pi \rho_d \kappa\left(a_{\max}-a_{\min}\right),\\
M_d &= \frac{4}{3}\rho_d C_xa_{\rm eff},
\end{split}
\end{equation}
Where $a_{\rm eff}$ is the effective particle radius
\begin{equation}
a_{\rm eff}
\equiv
\frac{a_{\max}-a_{\min}}
{\ln(a_{\max}/a_{\min})}.
\end{equation}
To convert this mass into a production rate, we assume that grains are lofted and move at some dust velocity $v_d$. The dust production rate is therefore
\begin{equation}
\dot{M}_d= M_d \frac{v_d}{\rho}.
\end{equation}
Combining the expressions for $C_x$ and $M_d$, we can finally write the dust production rate as
\begin{equation}
\dot{M}=\frac{4\pi}{3}\frac{a_{\rm eff}\rho_d v_d}{\varepsilon}\left(\varepsilon f \rho\right).
\end{equation}

\section{Laboratory Measurements of Spinel}
\label{sec:spinel}
The spinel (Mg\# = 0 and Cr\# = 99) shown in \autoref{fig:minerals} was originally sourced from Mogok, Myanmar, courtesy of G. Rossman's collection at the California Institute of Technology (GRR 2375). We performed mineralogy and composition verification using an INAM X-ray fluorescence (XRF) instrument with a 45 kV Ti source and silicon-drift solid-state detector. Measurements were taken in ambient conditions (without helium). After compositional characterization with the XRF, we ground the spinel with an agate mortar and pestle and sieved the powder into a 45–125-\um\ sized bin. The powder was not rinsed, thus clinging fines remain. MIR measurements were taken with a Thermo-Nicolet Fourier-Transform Infrared (FTIR) spectrometer under ambient conditions using a diffuse reflectance accessory, KBr beam splitter, KBr detector, and a 4 cm$^{-1}$ spectral resolution \citep[see][for additional measurement technique information]{Martin2022Icar}. To be directly comparable to the 3I emissivity spectra, the reflectances have been inverted via Kirchhoff's law.

\end{document}